\documentclass[superscriptaddress,prx,twocolumn,longbibliography]{revtex4-1}
\usepackage{graphicx}
\usepackage{amsmath,amssymb,physics,dsfont}
\usepackage{color}
\usepackage{hyperref}
\usepackage{microtype}
\usepackage{braket}
\usepackage{defs-private}
\usepackage{float}

\begin{document}

\title{Learning Spin Hamiltonians from Terahertz Two-Dimensional Coherent Spectroscopy}

\author{Martin Mootz}
\email{mootz@iastate.edu}
\affiliation{Ames National Laboratory, U.S. Department of Energy, Ames, Iowa 50011, USA}

\author{Chuankun Huang}
\affiliation{Ames National Laboratory, U.S. Department of Energy, Ames, Iowa 50011, USA}
\affiliation{Department of Physics and Astronomy, Iowa State University, Ames, Iowa 50011, USA}

\author{Liang Luo}
\affiliation{Ames National Laboratory, U.S. Department of Energy, Ames, Iowa 50011, USA}

\author{Jigang Wang}
\affiliation{Ames National Laboratory, U.S. Department of Energy, Ames, Iowa 50011, USA}
\affiliation{Department of Physics and Astronomy, Iowa State University, Ames, Iowa 50011, USA}

\author{Yong-Xin Yao}
\email{ykent@iastate.edu}
\affiliation{Ames National Laboratory, U.S. Department of Energy, Ames, Iowa 50011, USA}
\affiliation{Department of Physics and Astronomy, Iowa State University, Ames, Iowa 50011, USA}

\begin{abstract}
Effective Hamiltonians connect microscopic interactions to measurable collective behavior in quantum materials, but determining their parameters directly from experiment remains a challenging inverse problem. We introduce a supervised machine-learning framework that infers Hamiltonian parameters from nonlinear terahertz two-dimensional coherent spectra. A calibrated forward model generates spectra from candidate Hamiltonians, a common preprocessing pipeline maps simulated and experimental spectra into the same representation, and a neural network learns the inverse map from spectral fingerprints to microscopic parameters. We demonstrate the approach for rare-earth orthoferrites using a two-sublattice Landau--Lifshitz--Gilbert spin model with exchange, Dzyaloshinskii--Moriya interaction, anisotropies, and damping. Synthetic benchmarks show that nonlinear spectra encode parameters beyond those fixed by the linear response, with inference accuracy tracking the physical spectral sensitivity and robustness against noise improved by using multiple inter-pulse delays. Applied to experimental THz-2DCS data from Sm$_{0.4}$Er$_{0.6}$FeO$_3$, the inferred parameters yield physically reasonable forward simulations, while remaining discrepancies identify limitations of the reduced model. These results establish THz-2DCS as a data-rich platform for effective-Hamiltonian inference and model refinement, enabling experimentally driven identification of microscopic interactions while providing a foundation for understanding, predicting, and ultimately controlling the emergent properties of quantum materials.
\end{abstract}

\maketitle

\section{Introduction}

A central goal of quantum materials research is to identify the effective degrees of freedom and interactions that control measured collective behavior.
The full microscopic many-body Hamiltonian is generally too complex to connect directly to experiment, especially when electronic, spin, lattice, and electromagnetic degrees of freedom are intertwined. 
Experiments are therefore commonly interpreted in terms of effective Hamiltonians: tractable descriptions that retain the relevant low-energy variables while incorporating the influence of unresolved microscopic processes into material-specific parameters~\cite{Powell2009,auerbach1998interacting,Hubbard1963}. 
Depending on the material class, these parameters may describe hopping amplitudes and Coulomb interactions, electron--phonon or exciton--phonon couplings, superconducting pairing interactions, magnetic exchange and anisotropy terms, or light--matter coupling strengths. 
Determining such parameters from measurement is essential for turning spectroscopic data into microscopic understanding and for predicting equilibrium order, collective excitations, selection rules, and nonlinear nonequilibrium dynamics.

Reconstructing an effective Hamiltonian from experiment is, however, a difficult inverse problem. 
One usually starts from a candidate Hamiltonian whose form is motivated by symmetry, microscopic considerations, or prior measurements, and then seeks the parameter values that best reproduce the observed response. 
This strategy underlies conventional spin-wave analysis of neutron-scattering spectra~\cite{Toth_2015,Hahn:2014,Scheie2023NiPS3}, quantum-system-identification protocols based on measured time traces~\cite{Burgarth2012PRL,Zhang2014PRL}, and strong-field optical or THz approaches for reconstructing effective band or electron--hole Hamiltonians~\cite{Vampa2015PRL,Wu2026PRB,Borsch2020}. 
In many implementations, however, the comparison between model and experiment is reduced to a restricted set of observables, such as resonance frequencies, linewidths, dispersion relations, or the amplitudes of selected spectral peaks.
This approach is powerful when the relevant features are few and well separated, but it becomes fragile when distinct microscopic parameter sets produce nearly indistinguishable linear spectra. 
In such cases, the distinguishing information may be encoded in nonlinear spectral features, such as higher harmonics and field-induced mixing between collective modes, as well as in how these features depend on excitation and detection geometry. 
A central challenge is therefore to exploit these richer experimental response functions without reducing the data prematurely to a small number of manually chosen features.

\begin{figure}[t!]
\begin{center}
        \includegraphics[scale=0.48]{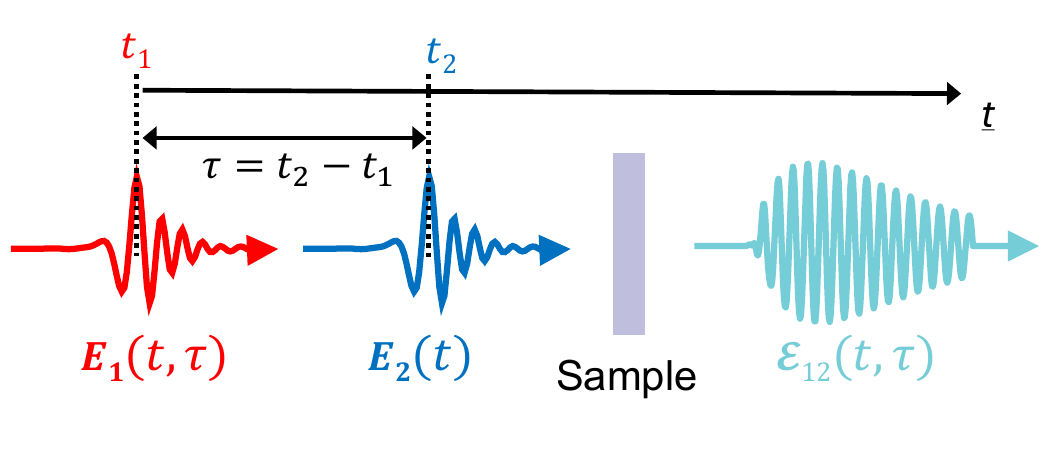}
        \caption{
Schematic of THz two-dimensional coherent spectroscopy in transmission geometry.
The sample is excited by two collinear, phase-locked few-cycle incident THz electric-field pulses. The first pulse, $\mathbf E_1$ (red), is centered at $t_1=-\tau$, while the second pulse, $\mathbf E_2$ (blue), is centered at $t_2=0$, defining the inter-pulse delay $\tau=t_2-t_1$. With this convention, positive $\tau$ corresponds to pulse 1 arriving before pulse 2. The field $\mathcal E_{12}(t,\tau)$ (cyan) denotes the transmitted electric field measured after excitation with both pulses present. The nonlinear THz response is obtained by subtracting the corresponding single-pulse transmitted reference fields, $\mathcal E_{\mathrm{NL}}(t,\tau)=\mathcal E_{12}(t,\tau)-\mathcal E_1(t,\tau)-\mathcal E_2(t)$. The resulting nonlinear response is resolved along both the detection-time axis $t$ and the inter-pulse-delay axis $\tau$.
}
        \label{fig1} 
\end{center}
\end{figure}

Terahertz two-dimensional coherent spectroscopy (THz-2DCS) provides a particularly rich nonlinear observable for Hamiltonian inference in driven quantum materials~\cite{Kuehn2009,Dutta2025rev,huang2026terahertz2d}. 
By using phase-locked THz pulse pairs (Fig.~\ref{fig1}), THz-2DCS coherently perturbs low-energy degrees of freedom and records the resulting phase-resolved nonlinear response as a function of both detection time and inter-pulse delay. 
In contrast to conventional one-dimensional spectroscopy, the resulting two-time response encodes excitation pathways, mode couplings, dephasing dynamics, and correlation effects that are difficult to isolate from linear spectra alone.
These multidimensional nonlinear spectra therefore provide high-dimensional fingerprints of how microscopic Hamiltonian terms shape driven dynamics, beyond the information contained in linear resonance frequencies and linewidths.

Multidimensional THz experiments have shown sensitivity to microscopic interactions across several classes of quantum materials. 
Early applications to semiconductors and lattice-coupled systems established THz-2DCS as a probe of intersubband dynamics, coherent phonon response, two-phonon quantum coherences, and anharmonic vibrational pathways~\cite{Kuehn2011,Somma2016PRL,Johnson2019,Blank2023Spin}. 
This capability has since been extended to superconducting systems, where experimental and theoretical studies of nonlinear THz spectroscopy and THz-2DCS have revealed and elucidated collective amplitude-mode dynamics, quasiparticle--collective-mode coupling, Higgs echoes, and Josephson plasmon echoes that are weak or hidden in linear response~\cite{Yang2018TerahertzQT,yang2019lightwave, mootz2022visualization,Vaswani2021,Higgs_2dTHz,Mootz2024,Katsumi2024NbN,huang2025discovery,Cheng2025,Liu2024,luo2026extreme}.
Magnetic materials provide another important example: intense THz pulses enable coherent control of antiferromagnetic spin waves~\cite{Kampfrath2011}, while multidimensional measurements have resolved specific manifestations of nonlinear spin dynamics, including magnon echoes, two-magnon coherences, magnon up- and down-conversion, and high-order magnon multiplication~\cite{Lu:2017,huang2024extreme,Zhang2024Up,Zhang2024Coup,Zhang2024Down}.
These examples show that THz-2DCS can access nonlinear fingerprints of interactions and couplings that are not uniquely constrained by linear spectra alone.

These experimental advances are complemented by theoretical studies showing how multidimensional nonlinear response can expose microscopic interaction effects, especially in spin systems where nonlinear susceptibilities can be related directly to Hamiltonian terms. 
Susceptibility-based and numerical studies have shown that 2DCS can probe fractionalization, confinement, interaction effects, spin-continuum structure, and disorder-induced spectral features in quantum spin liquids~\cite{Wan2019,Choi2020,Nandkishore2021,negahdariNonlinearResponseKitaev2023,qiang2023probing}, low-dimensional magnets~\cite{liPhotonEchoLensing2021,hartExtractingSpinonSelfenergies2023,gaoTwodimensionalCoherentSpectrum2023,simMicroscopicDetailsTwodimensional2023,liPhotonEchoFractional2023,Watanabe2024PRBSpinons,Zhang2024PRBContinua}, quantum spin ice~\cite{pottsExploitingPolarizationDependence2023,Watanabe2025PRLQuadrupolar}, and random magnets~\cite{parameswaranAsymptoticallyExactTheory2020}. 
More directly related to Hamiltonian inference, other work has identified fingerprints of specific microscopic terms and symmetries, including the magnitude and sign of the Dzyaloshinskii--Moriya interaction, anisotropy-induced nonlinear magnon response, and polarization-selective excitation pathways~\cite{ZhangTanimura2023JCP,mootz2023twodimensional,pottsExploitingPolarizationDependence2023}. 
Together, these developments suggest that THz-2DCS can encode microscopic information far beyond resonance frequencies alone. 
The missing ingredient is a robust inverse framework that transforms these nonlinear multidimensional spectral fingerprints into quantitative effective Hamiltonians, enabling direct interpretation of experiments in terms of the underlying microscopic physics.

Converting such nonlinear multidimensional fingerprints into Hamiltonian parameters requires solving a high-dimensional inverse problem, making learning-based approaches a natural complement to forward modeling. 
On the algorithmic side, Hamiltonian-learning protocols developed in quantum information and many-body theory have shown how local Hamiltonians can be reconstructed from local observables, eigenstates, or short-time dynamics~\cite{Bairey2019PRL,Qi2019Quantum,Yu2023robustefficient,Gu2024NatComm}. 
In parallel, materials-focused machine-learning strategies have been used to infer or refine effective microscopic models from complex experimental data, including neutron-scattering spectra~\cite{Samarakoon2020,Samarakoon2022CommunMater,Samarakoon2022PRR}, thermodynamic observables for crystal-field parameter estimation~\cite{Berthusen2021SciPost}, real-space magnetic images~\cite{Kwon2020,Wang2020AdvSci}, magnetometry~\cite{Fugetta2023PRR}, scanning tunneling microscopy~\cite{Sobral2023}, resonant inelastic x-ray scattering~\cite{RIXSInference2025}, and inelastic tunneling spectroscopy~\cite{Lupi2025PRApplied,Koch2025NanoLett}. 
These studies show that machine learning can be useful when a forward model is available and the measured observable contains high-dimensional, parameter-dependent structure that is difficult to reduce to a small set of manually selected features. 
This motivates a supervised-learning strategy for using the full frequency-resolved nonlinear THz response to infer Hamiltonian parameters, with forward simulations providing a direct validation of the inferred model.

Here we formulate Hamiltonian inference from THz-2DCS as a supervised learning problem. 
The approach combines a calibrated forward model, which maps microscopic Hamiltonian parameters and experimental inputs to nonlinear THz spectra, with a common preprocessing pipeline that maps simulated and experimental responses into the same spectral representation. 
A learned inverse model then maps these preprocessed nonlinear spectra back to Hamiltonian parameters. 
Rather than reducing the nonlinear response to a small set of manually selected quantities, such as peak positions, linewidths, or harmonic amplitudes, the network uses the frequency-domain spectra at selected inter-pulse delays as high-dimensional fingerprints of the underlying Hamiltonian. 
The inferred parameters are finally inserted back into the forward model, providing a direct validation step through comparison with the original nonlinear spectra and time-domain response.

We demonstrate this framework for spin-Hamiltonian inference in a weakly canted antiferromagnet motivated by THz measurements on rare-earth orthoferrite systems~\cite{huang2024extreme}. 
The forward model is an effective two-sublattice Landau--Lifshitz--Gilbert description with exchange, Dzyaloshinskii--Moriya interaction, quadratic and quartic anisotropies, and damping. 
For each sampled Hamiltonian parameter set, the exchange constant and Gilbert damping are calibrated to reproduce the experimentally relevant quasi-antiferromagnetic (qAFM) frequency and decay rate, while the remaining Hamiltonian parameters are inferred from the nonlinear response. 
Synthetic-data benchmarks are used to determine which parameters are encoded in different excitation and readout geometries, to quantify the role of spectral preprocessing and training-set size, and to test robustness against time-domain noise and the number of included inter-pulse delays. 
These benchmarks show that nonlinear spectra contain distinct fingerprints of the DM interaction and anisotropy terms, and that multi-delay inputs substantially improve the stability of the inference.

We then apply the developed Hamiltonian-inference framework to experimental THz-2DCS data from Sm$_{0.4}$Er$_{0.6}$FeO$_3$~\cite{huang2024extreme}. 
The experimental spectra are processed using the same preprocessing pipeline as the simulated data, and the inferred parameters are tested through forward simulations.
The resulting parameter sets are physically reasonable and reproduce part of the measured nonlinear spectral structure, demonstrating that experimental nonlinear THz spectra can be mapped onto effective Hamiltonian parameters within the assumed model class. 
At the same time, discrepancies in the third-harmonic spectral weight and delay-dependent time-domain phase reveal limitations of the effective two-sublattice model. 
Thus, the method provides both a route toward Hamiltonian reconstruction in driven magnetic materials and a diagnostic for identifying which ingredients are missing from the forward model.

This paper is organized as follows. Section~\ref{sec:framework} introduces THz-2DCS and formulates Hamiltonian inference as a supervised inverse problem. The framework is then specialized in Sec.~\ref{sec:forward_model} to an effective two-sublattice model for a weakly canted antiferromagnet motivated by rare-earth orthoferrites. Section~\ref{sec:dataset_ml} describes the generation of the synthetic training library, the spectral preprocessing pipeline, and neural-network-based parameter regression. Synthetic-data benchmarks in Sec.~\ref{sec:synthetic_results} quantify parameter sensitivity, polarization dependence, and how inference accuracy and noise robustness improve with the number of included inter-pulse delays. In Sec.~\ref{sec:experiment_application}, the trained inverse models are applied to experimental THz-2DCS data from Sm$_{0.4}$Er$_{0.6}$FeO$_3$ and the inferred parameters are tested through forward simulations. Finally, Sec.~\ref{sec:conclusion} summarizes the main results and discusses future directions toward quantitative Hamiltonian reconstruction from experimental THz-2DCS.

\section{General framework for Hamiltonian inference from THz-2DCS}
\label{sec:framework}

\subsection{Terahertz two-dimensional coherent spectroscopy}

Figure~\ref{fig1} illustrates a typical THz-2DCS experiment in a transmission geometry. Two collinear, phase-locked incident THz electric-field pulses, $\mathbf E_1$ and $\mathbf E_2$, excite the sample. We choose the center of the second pulse as the time reference, so that pulse 2 is centered at $t_2=0$ and pulse 1 is centered at $t_1=-\tau$. The inter-pulse delay is therefore $\tau=t_2-t_1$, such that positive $\tau$ corresponds to pulse 1 arriving before pulse 2.
For each value of $\tau$, the transmitted THz waveform $\mathcal E$ is recorded as a function of the detection time $t$ for excitation with both pulses present and for the corresponding single-pulse reference measurements.

We denote the relevant response channel generically by $S$. 
In experiment, $S$ corresponds to the transmitted electric-field waveform $\mathcal E$, including propagation and detection effects. 
In theory, $S$ may denote the corresponding simulated transmitted field or a microscopic response quantity such as magnetization, current, or polarization.
The nonlinear response is isolated by combining three measurements or simulations: the two-pulse response $S_{12}(t,\tau)$, obtained with both pulses applied, and the single-pulse reference responses $S_1(t,\tau)$ and $S_2(t)$. 
The nonlinear time-domain signal is then defined as
\begin{align}
S_{\mathrm{NL}}(t,\tau) = S_{12}(t,\tau) - S_1(t,\tau) - S_2(t)\,.
\label{eq:SNL_time}
\end{align}
For each fixed delay $\tau$, Fourier transformation with respect to the detection time $t$ gives
\begin{align}
S_{\mathrm{NL}}(\omega_t,\tau)
=
\mathcal{F}_t
\!\left[
S_{\mathrm{NL}}(t,\tau)
\right]\,.
\label{eq:SNL_freq}
\end{align}
The central goal is to use the information encoded in $S_{\mathrm{NL}}(\omega_t,\tau)$, for one or several selected delays, to infer the microscopic Hamiltonian parameters that govern the driven dynamics.

\begin{figure*}[t!]
\begin{center}
\includegraphics[scale=0.53]{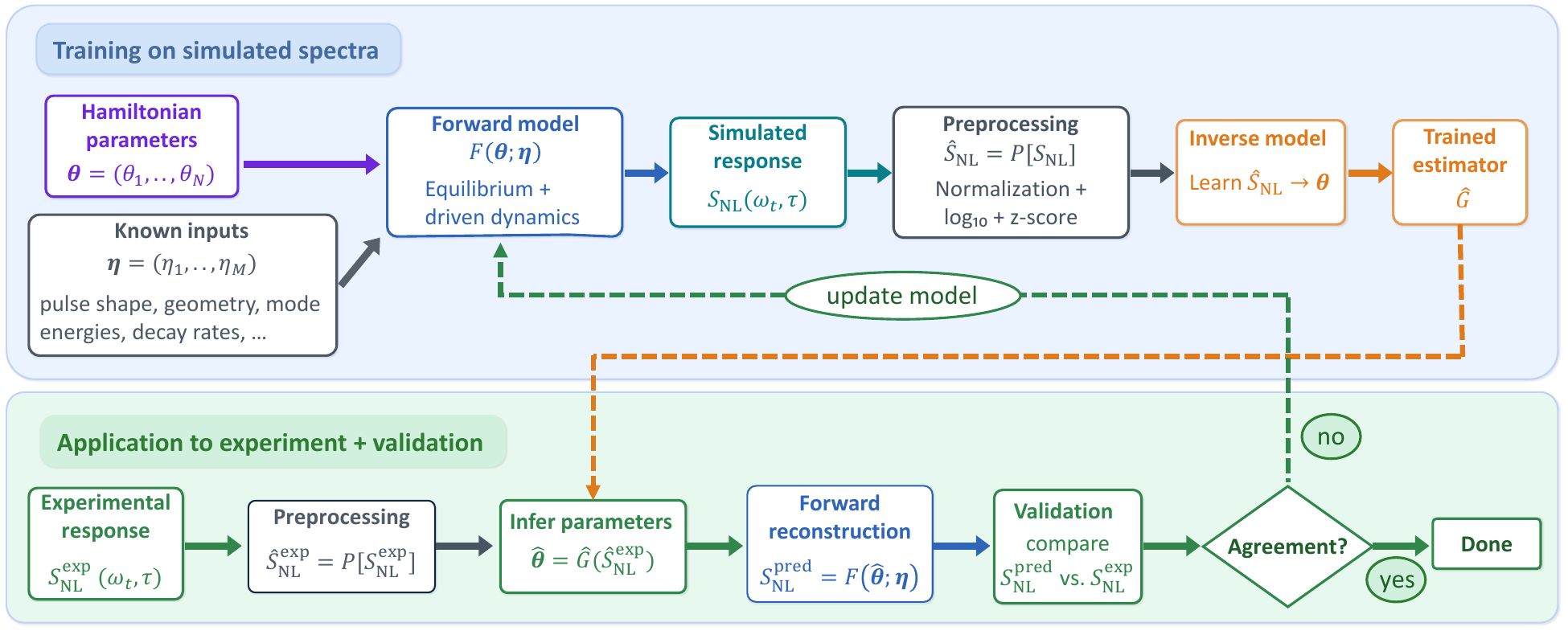}
\caption{
Workflow for machine-learning-based Hamiltonian inference from THz-2DCS.
The upper box shows the training stage based on simulated nonlinear spectra.  Hamiltonian parameters $\bth=(\theta_1,\ldots,\theta_N)$ are combined with known experimental and modeling inputs $\boldsymbol{\eta}=(\eta_1,\ldots,\eta_M)$, such as the pulse shapes, excitation geometry, decay rates, collective-mode energies, and analysis time window. These inputs define the forward model $F(\bth;\boldsymbol{\eta})$, which includes construction of the equilibrium state, propagation of the driven dynamics, and extraction of the nonlinear THz-2DCS response $S_{\mathrm{NL}}(\omega_t,\tau)$. The delay argument $\tau$ may denote either a single selected inter-pulse delay or a finite set of delays $\mathcal T=(\tau_1,\ldots,\tau_L)$, in which case the corresponding spectra are stacked as separate input channels. The simulated response is then transformed by the preprocessing map $P$, which may include normalization, logarithmic compression, and $z$-score standardization, to obtain the machine-learning input $\hat S_{\mathrm{NL}}=P[S_{\mathrm{NL}}]$. The preprocessed spectra and their corresponding Hamiltonian parameters are used to train an inverse regression model, resulting in the estimator $\hat G$.
The lower box shows the application to experiment and the validation loop. The measured nonlinear response $S_{\mathrm{NL}}^{\mathrm{exp}}(\omega_t,\tau)$ is processed using the same preprocessing map, $\hat S_{\mathrm{NL}}^{\mathrm{exp}}=P[S_{\mathrm{NL}}^{\mathrm{exp}}]$, and passed to the trained estimator to infer the experimental Hamiltonian parameters, $\hat{\bth}=\hat G[\hat S_{\mathrm{NL}}^{\mathrm{exp}}]$. These inferred parameters are inserted back into the forward model to generate a reconstructed nonlinear response $S_{\mathrm{NL}}^{\mathrm{pred}}=F(\hat{\bth};\boldsymbol{\eta})$, which is compared with the experimental response. Agreement supports the inferred parameters within the assumed forward model, while systematic discrepancies indicate that the Hamiltonian class, parameter ranges, observable mapping, preprocessing pipeline, or regression model may need to be refined.}
\label{fig2}
\end{center}
\end{figure*}

\subsection{Inverse problem and overall workflow}
\label{subsec:inverse_problem}

We formulate Hamiltonian inference from THz-2DCS as an inverse problem, as summarized in Fig.~\ref{fig2}. 
The Hamiltonian parameters to be inferred are denoted by $\bth=(\theta_1,\ldots,\theta_N)$, while $\boldsymbol{\eta}=(\eta_1,\ldots,\eta_M)$ denotes inputs fixed by the experimental configuration or by modeling choices.
These inputs include, for example, the pulse shape, excitation geometry, collective-mode energies, decay rates, detection window, and sampling conditions. 
For a given parameter set, the forward model maps the Hamiltonian parameters and known inputs to a nonlinear spectroscopic response,
\begin{align}
S_{\mathrm{NL}}(\omega_t,\tau)
=
F(\bth;\boldsymbol{\eta})\,.
\label{eq:general_forward_map}
\end{align}
Here $F$ denotes the complete simulation pipeline that converts a candidate Hamiltonian into the nonlinear THz-2DCS response. 
In general, this includes specifying the microscopic or effective Hamiltonian, determining the equilibrium state, applying the two-pulse THz driving protocol, propagating the driven dynamics, and extracting a nonlinear observable that can be compared with experiment. When several delays are used for learning, Eq.~\eqref{eq:general_forward_map} is evaluated at a finite set of selected delays, $\mathcal T=(\tau_1,\ldots,\tau_L)$, and the resulting spectra $\{S_{\mathrm{NL}}(\omega_t,\tau_\ell)\}_{\ell=1}^L$ are combined into one machine-learning input.

The inverse task is to reconstruct $\bth$ from the nonlinear response. Therefore, a library of simulated responses is generated by evaluating the forward model over a prescribed region of Hamiltonian-parameter space. 
Each simulated response is paired with the parameter vector used to generate it and treated as one supervised training example. 
Before training, the simulated spectra, either at a single delay or at a selected set of delays, are transformed by a preprocessing map,
\begin{align}
\hat S_{\mathrm{NL}}
=
P
\!\left[
S_{\mathrm{NL}}(\omega_t,\tau)
\right]\,,
\label{eq:preprocessing_map}
\end{align}
which may include normalization, logarithmic compression of the spectral dynamic range, and $z$-score standardization (converting data into units of standard deviations relative to the mean). 
The preprocessed spectra are then used to train an inverse regression model, $\hat G$, which approximates the map $\hat S_{\mathrm{NL}} \mapsto \bth$ from nonlinear THz-2DCS fingerprints to Hamiltonian parameters.

After training, the same preprocessing pipeline is applied to the experimental nonlinear response, using the same frequency grid, normalization convention, logarithmic compression, and training-set standardization parameters,
\begin{align}
\hat{S}_{\mathrm{NL}}^{\mathrm{exp}}
=
P
\!\left[
S_{\mathrm{NL}}^{\mathrm{exp}}(\omega_t,\tau)
\right]\,,
\label{eq:exp_preprocessing_map}
\end{align}
and the trained estimator is used to infer the experimental Hamiltonian parameters,
\begin{align}
\hat{\bth}
=
\hat{G}
\!\left[
\hat{S}_{\mathrm{NL}}^{\mathrm{exp}}(\omega_t,\tau)
\right]\,.
\label{eq:exp_inverse_map}
\end{align}
The inferred parameters are then inserted back into the forward model, together with the known experimental inputs $\boldsymbol{\eta}$, to generate a reconstructed nonlinear response,
\begin{align}
S_{\mathrm{NL}}^{\mathrm{pred}}(\omega_t,\tau)
=
F
\!\left(
\hat{\bth};
\boldsymbol{\eta}
\right)\,.
\label{eq:forward_reconstruction_map}
\end{align}
Comparison between $S_{\mathrm{NL}}^{\mathrm{pred}}$ and $S_{\mathrm{NL}}^{\mathrm{exp}}$ provides a consistency check for both the inferred parameters and the assumed forward model. If the agreement is poor, systematic discrepancies indicate that the Hamiltonian class, parameter ranges, observable mapping, preprocessing pipeline, or regression model may need to be refined.
This workflow emphasizes that THz-2DCS is not reduced to a small number of manually selected spectral features. Instead, the frequency-domain nonlinear response at selected delays is used as a high-dimensional fingerprint of the underlying Hamiltonian.

\begin{figure*}[t!]
\begin{center}
		\includegraphics[scale=0.52]{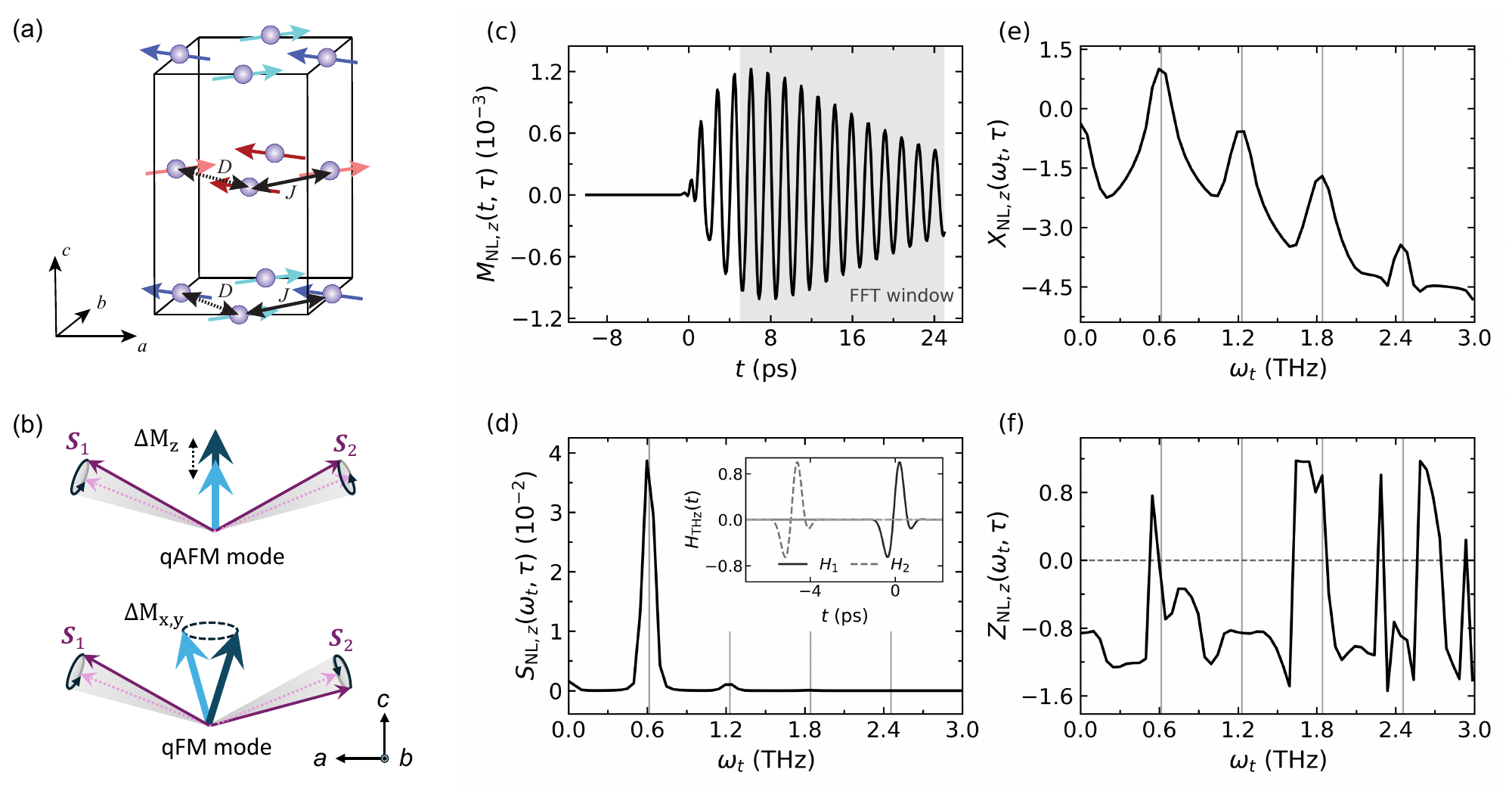}
\caption{
Spin-dynamics forward model and spectral preprocessing for machine learning.
(a) Schematic magnetic unit cell of the rare-earth orthoferrite structure motivating the model. The magnetic unit cell contains four Fe sublattices, with the colored arrows indicating the local Fe spin moments. The dominant nearest-neighbor exchange interaction $J$ and the Dzyaloshinskii--Moriya interaction $\mathbf D$ between neighboring Fe spin moments are indicated schematically. Exchange stabilizes the nearly antiferromagnetic arrangement, while the DM interaction, taken along the crystallographic $b$ axis, induces weak canting and produces a small net magnetization along the $c$ axis. The simulations reduce the four-sublattice magnetic structure to an effective two-sublattice model that captures the relevant zone-center spin dynamics.
(b) Schematic spin motion of the two low-energy collective modes. The quasi-ferromagnetic (qFM) mode corresponds mainly to a precession of the weak ferromagnetic moment, whereas the quasi-antiferromagnetic (qAFM) mode corresponds primarily to an oscillation of the canting angle, which appears as an amplitude modulation of the canting-induced net magnetization along the $c$ axis. These distinct spin motions lead to polarization-dependent coupling to
the THz magnetic field.
(c) Example nonlinear magnetization trace $M_{\mathrm{NL},z}(t,\tau)$ for a representative simulated parameter set at $\tau=4.8$ ps, with $z\parallel c$. The simulation uses $f_{\mathrm{qAFM}}=0.61~\mathrm{THz}$, $J=6.64~\mathrm{meV}$, $D=0.088~\mathrm{meV}$, $K_a=4.67\times10^{-3}~\mathrm{meV}$, and $K_c=K_4=0$. The shaded region marks the after-pulse detection window used for the spectral analysis.
(d) Frequency-domain amplitude spectrum $S_{\mathrm{NL},z}(\omega_t,\tau)$, obtained from the Fourier transform of the Hann-windowed nonlinear magnetization trace in panel (c), as defined in Eq.~\eqref{eq:preprocessing_fft}. The inset shows the corresponding two-pulse THz magnetic-field waveform.
(e) Log-compressed nonlinear spectrum
$X_{\mathrm{NL},z}(\omega_t,\tau)=1+\log_{10}[S_{\mathrm{NL},z}(\omega_t,\tau)/S_{\mathrm{NL},z}^{\mathrm{qAFM}}(\mathcal T)]$, where $\mathcal T=\{\tau\}$ for this single-delay example and $S_{\mathrm{NL},z}^{\mathrm{qAFM}}(\mathcal T)$ is the qAFM reference amplitude defined in the preprocessing procedure.
(f) Final standardized nonlinear input spectrum
$Z_{\mathrm{NL},z}(\omega_t,\tau)=[X_{\mathrm{NL},z}(\omega_t,\tau)-\mu_{\mathrm{NL},z}(\omega_t,\tau)]/
\sigma_{\mathrm{NL},z}(\omega_t,\tau)$, obtained using the frequency-dependent mean $\mu_{\mathrm{NL},z}(\omega_t,\tau)$ and standard deviation $\sigma_{\mathrm{NL},z}(\omega_t,\tau)$ computed from the training set. Vertical dashed lines in panels (d)--(f) mark integer harmonics of the fundamental qAFM resonance, $\omega_t=m f_{\mathrm{qAFM}}$.
}

		\label{fig3} 
\end{center}
\end{figure*}

\section{Forward model for a canted two-sublattice antiferromagnet}
\label{sec:forward_model}

We now specialize the general inference framework introduced in the previous section to a canted antiferromagnetic spin model motivated by THz-2DCS experiments on rare-earth orthoferrites~\cite{Lu:2017,huang2024extreme,Zhang2024Down,Zhang2024Coup,Zhang2024Up}. As a representative example, we focus on parameters relevant to Sm$_{0.4}$Er$_{0.6}$FeO$_3$, an orthorhombic rare-earth orthoferrite whose Fe moments form a weakly canted antiferromagnetic state below the N\'eel temperature~\cite{Zhao2016}. In this material, the Fe-spin structure undergoes a spin-reorientation transition between the $\Gamma_2$ and $\Gamma_4$ magnetic configurations over the temperature range $170$--$210$~K~\cite{Zhao2016}. In the present work, we study the room-temperature $\Gamma_4$ configuration, in which the antiferromagnetic vector is predominantly along the crystallographic $a$ axis and the weak ferromagnetic moment is along the $c$ axis. THz time-domain measurements in this phase show a pronounced quasi-antiferromagnetic (qAFM) resonance near $0.61$~THz for THz magnetic fields along $c$~\cite{Huang2024,Zhao2016}, providing a well-defined spectral feature for Hamiltonian inference from nonlinear THz-2DCS spectra.

Figure~\figref{fig3}{(a)} illustrates the magnetic unit cell of the rare-earth
orthoferrite structure. The unit cell contains four Fe sublattices, whose local spin moments (arrows) form a nearly antiferromagnetic arrangement. The dominant nearest-neighbor exchange interaction $J$ stabilizes this antiferromagnetic order, while the Dzyaloshinskii--Moriya (DM) interaction $\mathbf D$ produces a weak canting of the Fe moments. In the room-temperature $\Gamma_4$ configuration considered here, the dominant antiferromagnetic component lies along the crystallographic $a$ axis, while the canting produces a small net magnetization along the crystallographic $c$ axis.

Although the full orthorhombic magnetic unit cell contains four Fe sublattices, the low-energy zone-center dynamics relevant for the THz response can be represented by an effective two-sublattice classical spin model~\cite{Herrmann:1964}. In this reduced description, the four sublattice moments are represented by two effective collective classical spin vectors of fixed length $S$, denoted by $\mathbf S_1$ and $\mathbf S_2$. This effective model retains the dominant antiferromagnetic order, the weak DM-induced canting, and the anisotropy terms that, together with exchange and DM coupling, determine the low-energy collective-mode frequencies. The resulting equilibrium state is a weakly canted antiferromagnet with a small net magnetization along the crystallographic $c$ axis.

The two characteristic collective modes of this state are sketched in Fig.~\figref{fig3}{(b)}. The quasi-ferromagnetic (qFM) mode corresponds mainly to a precessional oscillation of the weak ferromagnetic moment around its equilibrium direction. The qAFM mode instead involves an oscillation of the canting angle between the two sublattice moments, which appears as an amplitude modulation of the canting-induced net magnetization along the crystallographic $c$ axis. Because these two modes involve different spin motions, their coupling to the THz magnetic field is strongly polarization dependent. For the room-temperature $\Gamma_4$ configuration considered here, THz magnetic fields oriented along the $c$ axis couple predominantly to the qAFM mode, whereas transverse fields, for example along the $a$ or $b$ axis, couple primarily to the qFM mode. The following subsections define the effective spin Hamiltonian, its equilibrium configuration, the two-pulse THz drive, and the nonlinear magnetization response used as the model-specific realization of $S_{\mathrm{NL}}$.

\subsection{Model Hamiltonian}
\label{subsec:model_hamiltonian}

Using the effective two-sublattice variables introduced above, we model the magnetic unit cell by the Hamiltonian
\begin{align}
\mathcal{H}
&=
n J\,\mathbf{S}_1\!\cdot\!\mathbf{S}_2
+
n\,\mathbf{D}\!\cdot\!\left(\mathbf{S}_1\times\mathbf{S}_2\right)
\nonumber \\ &\quad-
K_a\sum_{i=1,2} S_{ix}^2
-
K_c\sum_{i=1,2} S_{iz}^2
-
K_4\sum_{i=1,2}\sum_{\alpha=x,y,z} S_{i\alpha}^4
\nonumber\\
&\quad-
g\mu_B\sum_{i=1,2}\mathbf H_{\mathrm{THz}}(t,\tau)\!\cdot\!\mathbf S_i \,.
\label{eq:H_spin_model}
\end{align}
Throughout the model, the Cartesian spin components $(x,y,z)$ are identified with the crystallographic axes $(a,b,c)$, respectively. The vectors $\mathbf S_1$ and $\mathbf S_2$ are classical spins of fixed length $|\mathbf S_i|=S$, and we use $S=5/2$ for the Fe$^{3+}$ moments. 

The first term in Eq.~\eqref{eq:H_spin_model}, with antiferromagnetic exchange $J>0$, favors antiparallel alignment of the two sublattice spins and therefore stabilizes the dominant antiferromagnetic order. The factor $n$ accounts for the effective nearest-neighbor coordination in the two-sublattice model; in the simulations below we set $n=6$ according to the atomic configuration of the unit cell shown in Fig.~\figref{fig3}{(a)}.
The second term is the antisymmetric exchange term known as the Dzyaloshinskii--Moriya interaction. In the orthoferrite configuration considered here, the DM vector is taken along the crystallographic $b$ axis, $\mathbf{D}=D\hat{\mathbf y}$, with $\hat{\mathbf y}\parallel b$. This term favors a finite $b$-axis component of $\mathbf{S}_1\times\mathbf{S}_2$, and therefore stabilizes a weakly canted spin configuration in the $ac$ plane. In this configuration, the dominant $a$-axis components of the two sublattice moments are antiparallel, while their small $c$-axis components are parallel. The latter add to form the weak ferromagnetic moment along the crystallographic $c$ axis.

The terms proportional to $K_a$ and $K_c$ describe second-order magnetocrystalline anisotropy along the crystallographic $a$ and $c$ axes, respectively. With the convention $(x,y,z)\equiv(a,b,c)$ and the sign convention in Eq.~\eqref{eq:H_spin_model}, positive $K_a$ or $K_c$ lowers the energy when the corresponding spin component is large, thereby favoring spin components along that axis. For the room-temperature $\Gamma_4$ regime considered here, the relevant parameter region satisfies $K_a>K_c$, so that the quadratic anisotropy favors the dominant antiferromagnetic component along the $a$ axis over alignment along the $c$ axis. 
The quartic term proportional to $K_4$ introduces a higher-order onsite anisotropy, which modifies the angular dependence of the magnetic free energy beyond the quadratic approximation and is important for nonlinear spin dynamics under strong THz driving. 
The final term is the Zeeman coupling to the applied THz magnetic field and drives the nonequilibrium spin motion. We adopt the standard spin-dynamics convention in which the applied THz magnetic field $\mathbf H_{\mathrm{THz}}(t,\tau)$ is specified in tesla. Accordingly, the Zeeman interaction explicitly includes both the Bohr magneton $\mu_B$ and the Land\'e factor $g$ with $g=2$.

\subsection{Equilibrium configuration}
\label{subsec:equilibrium_state}

The equilibrium state used as the initial condition for the driven dynamics is a weakly canted antiferromagnet. Motivated by the room-temperature $\Gamma_4$ configuration of Sm$_{0.4}$Er$_{0.6}$FeO$_3$~\cite{Zhao2016}, we restrict the equilibrium spin configuration to the $ac$-plane form
\begin{align}
\mathbf{S}_1
&=
\bigl(S\sin\theta,0,S\cos\theta\bigr)\,,
\qquad
\mathbf{S}_2 = \bigl(-S\sin\theta,0, S\cos\theta\bigr)\,.
\label{eq:canted_ansatz}
\end{align}
In this parametrization, the dominant antiferromagnetic component lies along the crystallographic $a$ axis, while the parallel $c$-axis components produce the weak net magnetization. The angle $\theta$ therefore controls the canting of the two sublattice moments and fixes the magnitude of the equilibrium ferromagnetic component along $c$.

Substituting Eq.~\eqref{eq:canted_ansatz} into the static part of Eq.~\eqref{eq:H_spin_model} reduces the equilibrium problem to a single angular variable. Defining $\varphi=2\theta$ and dropping terms independent of $\varphi$, the energy can be expressed as
\begin{align}
E(\varphi) = S^2\left[A\cos\varphi - M\sin\varphi + K_4 S^2\sin^2\varphi \right]\,,
\label{eq:reduced_energy_phi}
\end{align}
with
\begin{align}
A=nJ+(K_a-K_c)\,,
\qquad
M=nD\,.
\label{eq:AM_def}
\end{align}

The equilibrium angle is obtained by solving the stationarity condition $\partial E/\partial\varphi=0$,
\begin{align}
-A\sin\varphi - M\cos\varphi + K_4S^2\sin(2\varphi) =0\,.
\label{eq:stationarity_phi}
\end{align}
This equation generally has multiple stationary solutions. The physical equilibrium angle $\varphi_{\mathrm{eq}}$ is chosen as the stationary point that minimizes the reduced energy $E(\varphi)$,
\begin{align}
\varphi_{\mathrm{eq}} = \arg\min_{\varphi\in\{\varphi:\,\partial E/\partial\varphi=0\}} E(\varphi)\,,
\qquad
\theta_{\mathrm{eq}}=\frac{\varphi_{\mathrm{eq}}}{2}\,.
\label{eq:phi_eq}
\end{align}
The corresponding spin configuration $\mathbf{S}_1(\theta_{\mathrm{eq}})$ and $\mathbf{S}_2(\theta_{\mathrm{eq}})$ is used as the initial condition for the time-dependent simulation. Technical details of the equilibrium-state calculation are given in Appendix~\ref{app:equilibrium}.

\subsection{Two-pulse driving, spin dynamics, and nonlinear observable}
\label{subsec:two_pulse_dynamics}

Starting from the equilibrium state constructed above, the system is driven by a pair of phase-locked THz magnetic-field pulses. We choose the center of pulse 2 as the time origin ($t_2=0$) and pulse 1 is centered at $t_1=-\tau$ (Fig.~\ref{fig1}). The total magnetic field is therefore written as
\begin{align}
\mathbf{H}_{\mathrm{THz}}(t,\tau)
=
\mathbf{H}_1(t+\tau)+\mathbf{H}_2(t)\,.
\label{eq:H_total}
\end{align}
Each pulse is modeled as a chirped Gaussian-envelope magnetic field,
\begin{align}
\mathbf{H}_j(t)
&=
H_j^{(0)}
\exp\!\left(-\frac{t^2}{\sigma_j^2}\right)
\sin\!\left[
2\pi f_j t\bigl(1+\beta_j t\bigr)+\phi_j
\right]\hat{\mathbf e}_j\,,
\nonumber\\
&j=1,2\,.
\label{eq:chirped_pulse}
\end{align}
Here $H_j^{(0)}$ is the pulse amplitude, $\sigma_j$ is the Gaussian envelope width, $f_j$ is the carrier frequency (central frequency), $\phi_j$ denotes the carrier-envelope phase, $\beta_j$ is the chirp parameter, and $\hat{\mathbf e}_j$ corresponds to the polarization direction of the magnetic field. 

The driven spin dynamics are propagated by solving the Gilbert-damped Landau--Lifshitz--Gilbert (LLG) equation~\cite{huang2024extreme},
\begin{align}
\frac{d\mathbf S_i}{dt}
=
-\frac{\gamma}{1+\alpha^2}
\left[
\mathbf S_i\times \mathbf H_i^{\mathrm{eff}}(t)
+
\frac{\alpha}{S}\,
\mathbf S_i\times
\bigl(
\mathbf S_i\times \mathbf H_i^{\mathrm{eff}}(t)
\bigr)
\right]\,,
\label{eq:LLG_model}
\end{align}
where the effective magnetic field is defined as
\begin{align}
\mathbf H_i^{\mathrm{eff}}(t)
=
-\frac{1}{g\mu_B}
\frac{\partial \mathcal H}{\partial \mathbf S_i}\,.
\label{eq:Heff_def}
\end{align}
Here $\gamma$ is the gyromagnetic ratio and $\alpha$ is the Gilbert damping parameter.

From the propagated trajectories we compute the net magnetization of the two-sublattice unit,
\begin{align}
\mathbf M(t)=\mathbf S_1(t)+\mathbf S_2(t)\,.
\label{eq:magnetization_def}
\end{align}
The nonlinear two-pulse response is then constructed in direct analogy to the experimental THz-2DCS protocol from three separate simulations: one with both pulses present, and two single-pulse reference calculations. This gives the nonlinear magnetization
\begin{align}
\mathbf M_{\mathrm{NL}}(t,\tau)
=
\mathbf M_{12}(t,\tau)-\mathbf M_1(t,\tau)-\mathbf M_2(t)\,,
\label{eq:Mnl_def_section}
\end{align}
where $\mathbf M_{12}$ denotes the response to the full pulse pair, while $\mathbf M_1$ and $\mathbf M_2$ are the corresponding single-pulse responses. The nonlinear magnetization component $M_{\mathrm{NL},\nu}(t,\tau)$, with $\nu\in\{x,y,z\}$, is the time-domain model observable from which the nonlinear spectrum $S_{\mathrm{NL},\nu}(\omega_t,\tau)$ is obtained by the Fourier transformation and preprocessing steps to be discussed in the next section.

\section{Dataset generation and machine-learning implementation}
\label{sec:dataset_ml}

The forward model described above provides a map from Hamiltonian parameters to a nonlinear THz-2DCS observable. We now describe how this map is used to generate synthetic training data, how the simulated nonlinear responses are converted into machine-learning inputs, and how the inverse regression model is trained. 

\subsection{Parameter sampling and calibration strategy}
\label{subsec:dataset_generation}

The synthetic training library is generated by sampling the Hamiltonian parameters to be inferred, $\bth=(D,K_a,K_c,K_4)$, and computing the corresponding nonlinear two-pulse response with the calibrated forward
model. The exchange constant $J$ and Gilbert damping parameter $\alpha$ are not treated as independent learning targets. Instead, they are fixed by calibration to the linear qAFM response: for each sampled $(D,K_a,K_c,K_4)$, $J$ is chosen such that the linearized qAFM mode reproduces the target resonance frequency specified below, and $\alpha$ is chosen such that the same mode reproduces the target decay rate. Thus, each nonlinear simulation is specified by $(D,K_a,K_c,K_4;J,\alpha)$, while only $(D,K_a,K_c,K_4)$ are used as supervised learning targets. The numerical procedures used to construct the equilibrium state and to calibrate $J$ and $\alpha$ are described in Appendix~\ref{app:forward_model}.

The independent Hamiltonian parameters are sampled from broad physically motivated intervals chosen to cover the energy scales commonly used for rare-earth orthoferrite spin models, while allowing for substantial variation around the representative material parameters,
\begin{align}
D &\in [0.02,0.165]~\mathrm{meV}\,, \nonumber \\
K_a &\in [5\times 10^{-4},10^{-2}]~\mathrm{meV}\,, \nonumber \\
K_c &\in [5\times 10^{-4},10^{-2}]~\mathrm{meV}\,, \nonumber \\
K_4 &\in [10^{-6},10^{-2}]~\mathrm{meV}\,.
\end{align}
All parameters are sampled uniformly within these intervals. We impose the condition $K_c<K_a$ to restrict the training set to the $\Gamma_4$-like regime relevant at room temperature, where the dominant antiferromagnetic component is along the crystallographic $a$ axis and the weak ferromagnetic moment is along $c$. In the sign convention of Eq.~\eqref{eq:H_spin_model}, this inequality makes the quadratic anisotropy favor the $a$-axis orientation over the $c$-axis orientation, consistent with the equilibrium configuration considered in Sec.~\ref{subsec:equilibrium_state}.

For the synthetic benchmarks in Sec.~\ref{sec:synthetic_results}, the calibration targets are chosen to represent the experimentally relevant qAFM response of the room-temperature orthoferrite. Specifically, we use a target qAFM frequency $f_{\mathrm{qAFM}}=0.61~\mathrm{THz}$ and decay rate $\Gamma_{\mathrm{qAFM}} = 0.077~\mathrm{ps}^{-1}$. These values define the linear frequency and damping scales used to calibrate $J$ and $\alpha$ for each sampled Hamiltonian parameter set. The use of a realistic finite decay rate is important for parameter inference from the nonlinear spectra. If the damping is too strong, coherent oscillations decay before higher-order nonlinear features build up, suppressing harmonics, sidebands, and line-shape distortions that are useful for parameter inference. The zero-damping limit is also not ideal because the response is purely coherent and long lived; after fixing the qAFM frequency and normalizing to the qAFM peak, the spectra can become dominated by sharp resonant features that provide a less robust fingerprint of the remaining Hamiltonian parameters. The experimentally motivated finite lifetime used here gives the response a realistic linewidth while preserving the nonlinear spectral features needed for learning.
Parameter sets are retained only if a physically acceptable calibrated Hamiltonian is found and the resulting equilibrium configuration satisfies the torque criterion described in Appendix~\ref{app:equilibrium}. Consequently, although the independent parameters are sampled uniformly at the proposal stage, the distribution of accepted parameter sets entering the machine-learning library is not strictly uniform over the original sampling box.

For the default fully synthetic benchmark datasets, each library contains $3.1\times10^4$ accepted parameter sets out of a total of 38547 points on a uniform parameter mesh. In these datasets, the two THz magnetic-field pulses are taken to be identical apart from their relative delay: they are collinear, have the same envelope and carrier parameters, $\sigma=0.55~\mathrm{ps}$, $f=0.68~\mathrm{THz}$, $\phi=0.42$, and $\beta=0.2~\mathrm{ps}^{-1}$, and have equal amplitudes $H_1^{(0)}=H_2^{(0)}\equiv H_{\mathrm{pulse}}$. For the $z$-polarized benchmarks, we use $\hat{\mathbf e}_1=\hat{\mathbf e}_2=\hat{\mathbf z}$. For the mixed-polarization benchmarks, we use $\hat{\mathbf e}_1=\hat{\mathbf e}_2=(\hat{\mathbf y}+\hat{\mathbf z})/\sqrt{2}$. We consider the pulse amplitudes $H_{\mathrm{pulse}}=0.1,0.5,1.0,2.0,4.0~\mathrm{T}$. Single-delay datasets use $\tau=4.8~\mathrm{ps}$, while multi-delay datasets are generated by evaluating the same calibrated parameter library for several values of $\tau$. For each accepted parameter set and pulse configuration, the LLG equations in Eq.~\eqref{eq:LLG_model} are propagated over the time interval $t\in[-10,25]~\mathrm{ps}$ using a fourth-order Runge--Kutta method with time step $\Delta t=10^{-2}~\mathrm{ps}$.

This idealized equal-pulse parametrization is used only for the fully synthetic benchmark study. For the experimental inference in Sec.~\ref{sec:experiment_application}, the two THz pulse waveforms are instead specified separately from fits to the measured THz fields, and the linear qAFM frequency and decay rate used for calibration are obtained from fits to the transmitted qAFM response, as described in Appendix~\ref{app:experimental_fits}.

\subsection{Spectral representation and preprocessing}
\label{subsec:preprocessing}

Figures~\figref{fig3}{(c-f)} illustrate how a simulated nonlinear magnetization trace is converted into the spectral input used for machine learning. The example shown is the $z$-axis nonlinear magnetization $M_{\mathrm{NL},z}(t,\tau)$ for a representative delay and parameter set. The same preprocessing steps are applied to any magnetization component $\nu\in\{x,y,z\}$ and to each selected inter-pulse delay $\tau$. 

For each delay $\tau$, we first remove the static pre-pulse background of the nonlinear magnetization trace. This step is needed because the LLG simulations output the absolute magnetization, so each pre-pulse trace contains the same equilibrium magnetization $M_{0,\nu}$. Consequently, the nonlinear subtraction $M_{12,\nu}-M_{1,\nu}-M_{2,\nu}$ leaves a static pre-pulse offset $-M_{0,\nu}$. We therefore define the baseline-corrected nonlinear magnetization as
\begin{align}
\widetilde M_{\mathrm{NL},\nu}(t,\tau)
&=
M_{12,\nu}(t,\tau)
-
M_{1,\nu}(t,\tau)
-
M_{2,\nu}(t)
+
M_{0,\nu}
\nonumber\\
&=
M_{\mathrm{NL},\nu}(t,\tau)
-
M_{\mathrm{NL},\nu}(t_{\mathrm{ref}},\tau)\,,
\label{eq:Mnl_equilibrium_subtraction}
\end{align}
where $t_{\mathrm{ref}}=-10$~ps is the initial simulation time before the arrival of the THz pulses. 
This sets the pre-pulse background to zero and isolates the driven nonlinear dynamics. Figure~\figref{fig3}{(c)} shows the resulting time-domain nonlinear response for the representative $z$-axis example. The shaded region marks the after-pulse detection window used for the spectral analysis. For the synthetic datasets used below, this window is $t\in[5,25]~\mathrm{ps}$. This choice mirrors the analysis of THz-2DCS experiments, where restricting the Fourier transform to an after-pulse detection window emphasizes the freely evolving dynamics after the excitation pulses have passed and reduces sensitivity to pulse-overlap contributions or non-resonant artifacts during the excitation window~\cite{huang2025discovery}.

Before Fourier transformation, the selected baseline-corrected time-domain signal is multiplied by a Hann window to reduce artifacts caused by the finite detection window. For a detection window containing $N_t$ sampled time points $t_m$, the windowed nonlinear magnetization component is
\begin{align}
M^{\mathrm{win}}_{\mathrm{NL},\nu}(t_m,\tau)
&=
w_m \widetilde M_{\mathrm{NL},\nu}(t_m,\tau)\,, \nonumber \\
w_m
&=
\frac{1}{2}
\left[
1-\cos\left(\frac{2\pi m}{N_t-1}\right)
\right]\,,
\end{align}
with $m=0,\ldots,N_t-1$. For each fixed inter-pulse delay $\tau$, we then compute the magnitude of the Fourier transform with respect to detection time,
\begin{align}
S_{\mathrm{NL},\nu}(\omega_t,\tau)
=
\left|
\mathcal{F}_{t\rightarrow \omega_t}
\left[
M^{\mathrm{win}}_{\mathrm{NL},\nu}(t,\tau)
\right]
\right|\,.
\label{eq:preprocessing_fft}
\end{align}
In the numerical implementation, $\mathcal{F}_{t\rightarrow\omega_t}$ is evaluated by a discrete FFT over the sampled times $t_m$.
Figure~\figref{fig3}{(d)} shows the resulting Fourier amplitude for the same representative trace. The spectrum contains pronounced nonlinear harmonic features, with peaks appearing at integer multiples of the fundamental qAFM frequency, as indicated by the vertical lines. For the learning inputs used below, we retain the frequency range $0\leq \omega_t \leq 2.7~\mathrm{THz}$, which contains the fundamental qAFM response and the relevant higher-harmonic features.

The Fourier amplitudes contain both relative spectral-shape information (e.g., peak ratios and linewidths) and an overall amplitude scale. The latter is not directly comparable between the simulated nonlinear magnetization $M_{\mathrm{NL},\nu}$ and the experimentally measured transmitted THz electric field $\mathcal E_{\mathrm{NL},\nu}$, because their relation depends on electromagnetic propagation through the sample, the detection response, and possible calibration factors. Moreover, the fundamental magnetic resonance is typically much stronger than the higher-harmonic and weaker nonlinear features. We therefore normalize the nonlinear spectra by a common qAFM reference amplitude. For a selected set of inter-pulse delays $\mathcal T$, we define
\begin{align}
S^{\mathrm{qAFM}}_{\mathrm{NL},\nu}(\mathcal T)
=
\max_{\tau\in\mathcal T}
\max_{\omega_t\in\mathcal W_{\mathrm{qAFM}}}
S_{\mathrm{NL},\nu}(\omega_t,\tau)\,,
\end{align}
where $\mathcal W_{\mathrm{qAFM}}$ is a narrow frequency window around
$f_{\mathrm{qAFM}}\simeq0.61~\mathrm{THz}$. Thus, when several delays are included in the machine-learning input, all corresponding spectra are normalized by the largest qAFM peak amplitude among those delays. For a single-delay input, this definition reduces to normalization by the qAFM peak amplitude of that individual spectrum. This normalization preserves the relative spectral structure and signal amplitudes between the included delays while removing the overall absolute signal scale, which is not used as a learning target in the present implementation. The normalized nonlinear spectrum is then logarithmically compressed and shifted by one,
\begin{align}
X_{\mathrm{NL},\nu}(\omega_t,\tau)
=
1+\log_{10}
\left[
\max\left(
\frac{S_{\mathrm{NL},\nu}(\omega_t,\tau)}
{S^{\mathrm{qAFM}}_{\mathrm{NL},\nu}(\mathcal T)},
\epsilon
\right)
\right]\,,
\label{eq:preprocessing_log}
\end{align}
where $\epsilon=10^{-12}$ regularizes the logarithm near zero. With this convention, the largest fundamental qAFM peak among the included delays has value $X_{\mathrm{NL},\nu}=1$, while features one order of magnitude weaker have $X_{\mathrm{NL},\nu}=0$. This log-compressed nonlinear spectrum is shown in Fig.~\figref{fig3}{(e)}. The normalization expresses the nonlinear response relative to the fundamental qAFM resonance, while the logarithmic scaling enhances the visibility of weaker harmonic features.

Finally, the log-compressed nonlinear spectra are standardized using the mean and standard deviation computed from the training set only. For a training set containing $N_{\mathrm{train}}$ spectra, these quantities are defined on a frequency-by-frequency and delay-by-delay basis as
\begin{align}
&\mu_{\mathrm{NL},\nu}(\omega_t,\tau)
=
\frac{1}{N_{\mathrm{train}}}
\sum_{r=1}^{N_{\mathrm{train}}}
X_{\mathrm{NL},\nu}^{(r)}(\omega_t,\tau)\,,
\nonumber\\
&\sigma_{\mathrm{NL},\nu}(\omega_t,\tau) \nonumber \\
&=
\left[
\frac{1}{N_{\mathrm{train}}}
\sum_{r=1}^{N_{\mathrm{train}}}
\left(
X_{\mathrm{NL},\nu}^{(r)}(\omega_t,\tau)
-
\mu_{\mathrm{NL},\nu}(\omega_t,\tau)
\right)^2
\right]^{1/2}\,,
\end{align}
where $r$ labels training samples. The standardized nonlinear input is then
\begin{align}
Z_{\mathrm{NL},\nu}(\omega_t,\tau)
=
\frac{
X_{\mathrm{NL},\nu}(\omega_t,\tau)-\mu_{\mathrm{NL},\nu}(\omega_t,\tau)
}{
\sigma_{\mathrm{NL},\nu}(\omega_t,\tau)
}\,.
\label{eq:preprocessing_zscore}
\end{align}
The same training-set values of $\mu_{\mathrm{NL},\nu}$ and $\sigma_{\mathrm{NL},\nu}$ are used for validation, testing, and experimental inference so that all inputs are represented in the same standardized feature space learned during training. This also avoids information leakage from the validation or test sets into the preprocessing. Figure~\figref{fig3}{(f)} shows the final standardized nonlinear spectrum used as the machine-learning input. For a single-delay, single-component dataset, this spectrum forms a one-dimensional input vector. For multi-delay or multi-component datasets, the corresponding standardized spectra are stacked along the channel dimension.

\begin{figure*}[t!]
\begin{center}
		\includegraphics[scale=0.4]{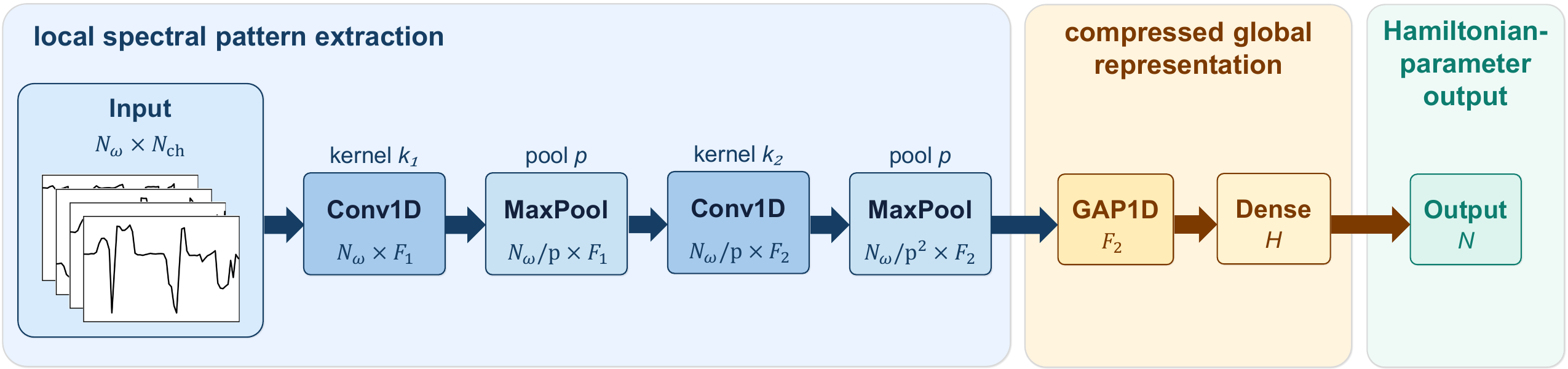}
\caption{
One-dimensional CNN architecture for Hamiltonian inference.
The input is a standardized nonlinear spectrum represented as an array of size
$N_\omega\times N_{\mathrm{ch}}$, where $N_\omega$ is the number of retained frequency points and $N_{\mathrm{ch}}$ is the number of spectral channels. For single-delay, single-component inputs, $N_{\mathrm{ch}}=1$; for multi-delay or multi-component inputs, the corresponding standardized spectra are stacked along the channel dimension. Two Conv1D--MaxPool blocks extract local spectral patterns along the frequency axis. The kernel sizes $k_1$ and $k_2$ specify the number of neighboring spectral points used by each learned filter, with the second convolution using dilation to enlarge the effective receptive range. Global average pooling compresses the resulting feature map into a compact representation, which is passed through a dense hidden layer of width $H$ and a final regression head to predict the $N$ Hamiltonian parameters.
}
		\label{fig4} 
\end{center}
\end{figure*}

\subsection{Neural-network architecture and training protocol}
\label{subsec:ml_architecture}

The standardized nonlinear spectra are used as input to a one-dimensional convolutional neural network (CNN), illustrated schematically in Fig.~\ref{fig4}. The input is represented as
$\mathbf Z \in \mathbb{R}^{N_\omega \times N_{\mathrm{ch}}}$,
where $N_\omega$ is the number of retained frequency points and $N_{\mathrm{ch}}$ is the number of input channels. For a single delay and a single magnetization component, $N_{\mathrm{ch}}=1$. For multi-delay datasets, the standardized spectra at different values of $\tau$ are stacked along the channel dimension. If several magnetization components are included, the channel dimension similarly combines delay and component indices. Thus, the same architecture can be used for single-delay, multi-delay, and multi-component inputs.

The CNN contains two convolutional blocks followed by global average pooling and a dense regression head. Starting from the input array
$\mathbf Z\in\mathbb R^{N_\omega\times N_{\mathrm{ch}}}$, the first convolutional layer uses $F_1=64$ filters with kernel size $k_1=7$ and maps the input to a frequency-resolved feature map of size $N_\omega\times F_1$, with zero-padding applied at the boundaries. The first max-pooling layer, with pool size $p=2$, reduces this feature map to size $(N_\omega/p)\times F_1$. The second convolutional layer uses $F_2=128$ filters with kernel size $k_2=5$ and dilation rate $d=2$, producing a feature map of size $(N_\omega/p)\times F_2$. The second max-pooling layer then reduces the feature map to size $(N_\omega/p^2)\times F_2$. The convolutional kernels act along the frequency axis and learn local spectral patterns such as resonance peaks, harmonic features, and line-shape changes. The dilation in the second convolution increases the effective receptive range relative to a standard adjacent five-point kernel.

After the two convolutional blocks, global average pooling averages over the remaining frequency dimension and produces a fixed $F_2$-dimensional feature vector. This step summarizes the frequency-resolved feature map before the dense regression head. The feature vector is passed through a dense hidden layer with $H=128$ neurons and then through a final dense output layer with $N$ neurons, where $N$ is the number of Hamiltonian parameters to be inferred. Rectified-linear activations are used in the convolutional and hidden dense layers, while the output layer uses a sigmoid activation. Accordingly, each target Hamiltonian parameter is mapped to a dimensionless value in the interval $[0,1]$ during training using a min--max transformation,
\begin{align}
\theta_j^{\mathrm{scaled}}
=
\frac{\theta_j-\theta_j^{\min}}
{\theta_j^{\max}-\theta_j^{\min}}\,,
\qquad j=1,\ldots,N\,,
\label{eq:minmax_target_scaling}
\end{align}
where $\theta_j^{\min}$ and $\theta_j^{\max}$ are the lower and upper bounds of the corresponding sampled parameter range. After prediction, the inverse transformation is applied to recover the Hamiltonian parameters in physical units. This target scaling is distinct from the $z$-score standardization applied to the spectral input.

The model is trained in a supervised manner to learn the mapping from standardized nonlinear spectra to the corresponding Hamiltonian parameters. For each dataset size considered, $N_{\mathrm{samples}}$ simulated spectra are used for model development and are split into training and validation subsets. The training subset is used to compute the preprocessing statistics and to update the network weights, whereas the validation subset is used only to monitor training and select the final model. An independent test set is kept separate from $N_{\mathrm{samples}}$ and is used only after training to evaluate the final prediction accuracy. Unless otherwise stated, 20\% of the $N_{\mathrm{samples}}$ model-development spectra are used for validation, the network is trained for 500 epochs with a batch size of 32, and the Adam optimizer is used with learning rate $3\times10^{-4}$. The loss function minimized during training is the mean-squared error between the predicted and true scaled Hamiltonian parameters~\eqref{eq:minmax_target_scaling}. The mean absolute error of the scaled parameters is recorded as an additional diagnostic during training but is not used to update the network weights.

To quantify the accuracy of the inverse-model predictions and assess the dependence on dataset size and statistical variability, the training is repeated for different values of $N_{\mathrm{samples}}$ and different random seeds. Prediction errors are evaluated parameter by parameter on an independent test set of size $N_{\mathrm{test}}$. Since the Hamiltonian parameters have different absolute scales, we use the relative mean absolute error,
\begin{align}
\mathrm{rel.\ MAE}_j
=
\frac{1}{N_{\mathrm{test}}}
\sum_{m=1}^{N_{\mathrm{test}}}
\frac{
\left|
\hat{\theta}_{j}^{(m)}-\theta_{j}^{(m)}
\right|
}{
\left|\theta_{j}^{(m)}\right|
}\,,
\end{align}
and the relative maximum error,
\begin{align}
\mathrm{rel.\ max}_j
=
\max_{m}
\frac{
\left|
\hat{\theta}_{j}^{(m)}-\theta_{j}^{(m)}
\right|
}{
\left|\theta_{j}^{(m)}\right|
}\,.
\end{align}
Here $\theta_j^{(m)}$ and $\hat{\theta}_j^{(m)}$ are the true and predicted values of parameter $\theta_j$ for test sample $m$, respectively. The relative MAE measures the average relative prediction error, while the relative maximum error captures the worst-case relative error in the test set. The inferred parameters are then used for forward validation by comparing the reconstructed nonlinear response with the reference spectrum.

\begin{figure*}[t!]
\begin{center}
\includegraphics[scale=0.50]{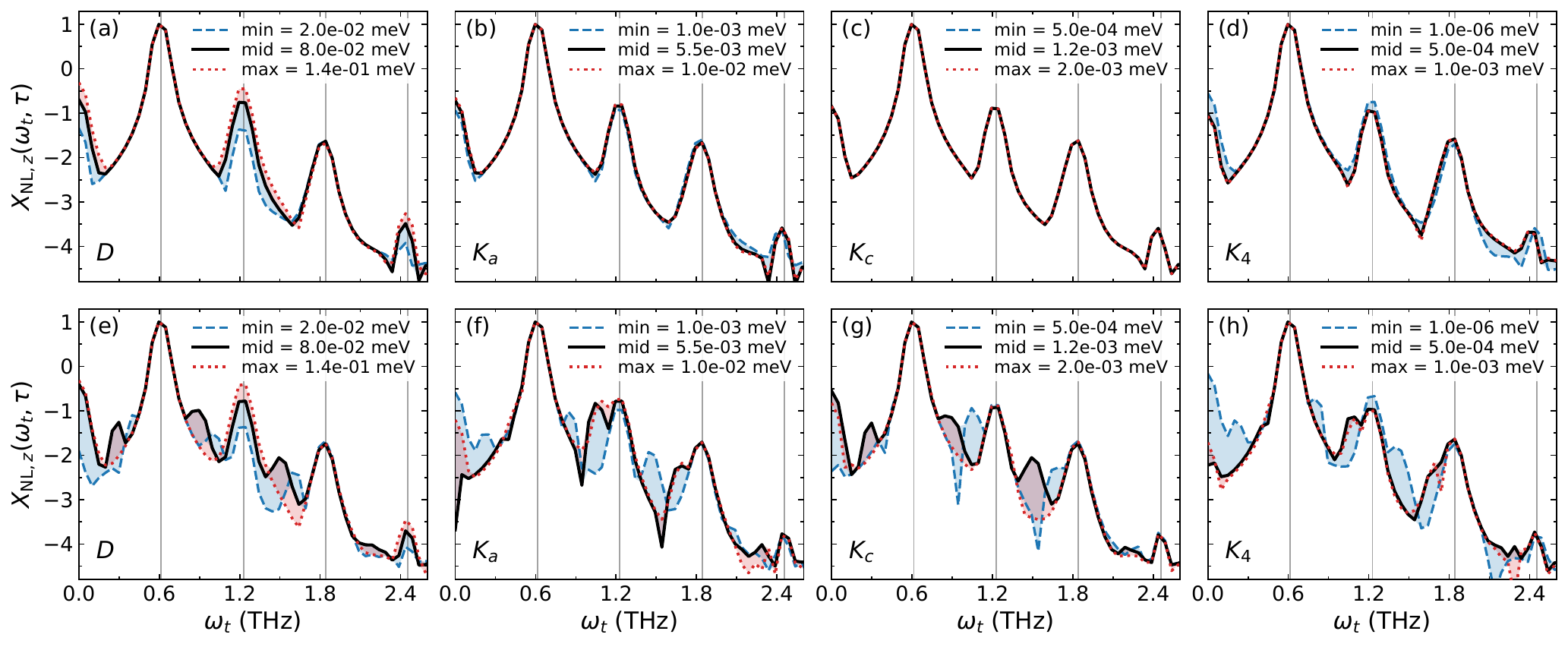}
\caption{
Parameter sensitivity of synthetic nonlinear spectra.
One-at-a-time parameter sweeps for two pulse configurations. In all panels, the plotted signal is the $z$-component nonlinear magnetization spectrum. 
The plotted quantity is the log-compressed normalized nonlinear spectrum $X_{\mathrm{NL},z}(\omega_t,\tau)$ defined in Eq.~\eqref{eq:preprocessing_log}, evaluated at $\tau=4.8~\mathrm{ps}$ for equal pulse amplitudes $H_1^{(0)}=H_2^{(0)}=1~\mathrm{T}$.
Panels (a)--(d) correspond to $z$-polarized excitation, $\hat{\mathbf e}_1=\hat{\mathbf e}_2=\hat{\mathbf z}$, while panels (e)--(h) show equal $y$--$z$-polarized excitation, $\hat{\mathbf{e}}_1=\hat{\mathbf{e}}_2=(\hat{\mathbf{y}}+\hat{\mathbf{z}})/\sqrt{2}$; in this case the $x$ and $y$ response components are also finite, but only the $z$ component is shown here for direct comparison with panels (a)--(d).
Columns show sweeps of (a,e) $D$, (b,f) $K_a$, (c,g) $K_c$, and (d,h) $K_4$. In each sweep, one Hamiltonian parameter is varied while the remaining parameters are fixed at $D=0.06~\mathrm{meV}$, $K_a=2.0\times 10^{-3}~\mathrm{meV}$, $K_c=1.0\times 10^{-3}~\mathrm{meV}$, and $K_4=10^{-4}~\mathrm{meV}$.
For each parameter value, $J$ is recalibrated to keep the qAFM frequency fixed, and $\alpha$ is recalibrated to match the target qAFM decay rate. In each panel, the curves correspond to the minimum (dashed line), midpoint (solid line), and maximum value (dotted line) of the swept parameter; shaded regions highlight the spectral changes between these representative values.
Vertical guide lines mark integer multiples of the fundamental qAFM frequency.
}
\label{fig5}
\end{center}
\end{figure*}

\section{Benchmarking on synthetic data}
\label{sec:synthetic_results}

Before applying the inverse model to experimental data, we benchmark the full workflow on synthetic datasets for which the underlying Hamiltonian parameters are known. This controlled setting allows us to separate several aspects of the inverse problem: the physical sensitivity of nonlinear spectra to individual Hamiltonian terms, the dependence of inference accuracy on excitation polarization and measured magnetization component, the role of preprocessing and training-library size, the learnability of higher-order anisotropy terms, and the robustness of the inferred parameters to time-domain noise. We first analyze one-at-a-time parameter sweeps to identify the nonlinear spectral fingerprints of the Hamiltonian parameters. We then quantify CNN-based parameter inference for several synthetic benchmark tasks and use forward simulations with the inferred parameters to test whether the reconstructed nonlinear dynamics reproduce the reference response.

\subsection{Parameter sensitivity of synthetic nonlinear spectra}
\label{subsec:parameter_sensitivity}

Before training the neural network, we first examine how the nonlinear spectra respond to individual Hamiltonian parameters. This provides a physical interpretation of the information available to the learning model and helps explain why some parameters are easier to infer than others. Figure~\ref{fig5} shows one-at-a-time parameter sweeps for two pulse configurations. All panels show the $z$-component log-compressed normalized nonlinear spectrum
$X_{\mathrm{NL},z}(\omega_t,\tau)$ at $\tau=4.8~\mathrm{ps}$. In each sweep, one parameter is varied while the remaining parameters are kept fixed at $D=0.06~\mathrm{meV}$, $K_a=2.0\times10^{-3}~\mathrm{meV}$, $K_c=1.0\times10^{-3}~\mathrm{meV}$, and $K_4=10^{-4}~\mathrm{meV}$. For every parameter value, the exchange coupling $J$ is recalibrated so that the qAFM frequency remains fixed, and the Gilbert damping parameter $\alpha$ is recalibrated so that the qAFM decay rate stays at its target value. Thus, the spectral changes in Fig.~\ref{fig5} reflect changes in the nonlinear response at fixed linear qAFM resonance frequency and decay rate, rather than trivial shifts or broadening changes of the dominant linear mode.

We first discuss the $z$-polarized excitation, shown in Figs.~\figref{fig5}{(a-d)}. In this geometry, the nonlinear net magnetization remains restricted to the $z$-axis response channel in the reduced two-sublattice model: $M_{\mathrm{NL},x}(t,\tau)=M_{\mathrm{NL},y}(t,\tau)=0$. Therefore, the parameter sensitivity is determined by how each Hamiltonian term affects $M_{\mathrm{NL},z}(t,\tau)$. The $D$ sweep, Fig.~\figref{fig5}{(a)}, strongly modifies the even-harmonic response, consistent with the role of the DM interaction in controlling weak canting and inversion-symmetry breaking of the magnetic structure. The $K_a$ sweep, Fig.~\figref{fig5}{(b)}, produces more subtle line-shape changes and peak asymmetries, indicating that $K_a$ is harder to infer from $z$-polarized spectra than $D$.

By contrast, the $K_c$ sweep, Fig.~\figref{fig5}{(c)}, produces no visible change in the spectrum in this geometry. This can be understood from the $K_c$ contribution to the $z$-component equation of motion. The $K_c$ anisotropy generates the effective field
\begin{align}
\mathbf H_i^{(K_c)}
=
\frac{2K_c}{g\mu_B}S_{iz}\hat{\mathbf z}\,.
\end{align}
The precessional part of the LLG equation has no $z$ component from this anisotropy field,
\begin{align}
\left[
\mathbf S_i\times \mathbf H_i^{(K_c)}
\right]_z=0\,.
\end{align}
For finite Gilbert damping, the same anisotropy contributes to the $z$-component equation only through the damping term,
\begin{align}
\left.
\frac{dS_{iz}}{dt}
\right|_{K_c}
=
\frac{2\gamma\alpha K_c}
{g\mu_B S(1+\alpha^2)}
S_{iz}\left(S_{ix}^2+S_{iy}^2\right)\,.
\end{align}
Thus, $K_c$ does not enter the leading precessional dynamics of $M_{\mathrm{NL},z}$ for $z$-polarized excitation. Its finite-damping contribution is proportional to $\alpha K_c$ and is therefore much smaller than the direct precessional contributions from $D$, $K_a$, and $K_4$. Consequently, within the parameter range and damping values used here, $K_c$ does not generate an independent visible spectral fingerprint in the normalized $z$-component spectra.
Finally, the $K_4$ sweep, Fig.~\figref{fig5}{(d)}, changes both peak heights and peak asymmetries, reflecting the role of the quartic anisotropy in the nonlinear spin dynamics under strong THz driving.

We next discuss the equal $y$--$z$-polarized excitation,
$\hat{\mathbf e}_1=\hat{\mathbf e}_2=(\hat{\mathbf y}+\hat{\mathbf z})/\sqrt{2}$, shown in
Figs.~\figref{fig5}{(e-h)}. In this geometry, all four Hamiltonian parameters modify the nonlinear spectra. The additional $y$-polarized magnetic-field component excites the qFM mode in addition to the qAFM mode, producing sideband-like spectral structures around the qAFM harmonic features. As shown in Figs.~\figref{fig5}{(e,f)}, variations of $D$ and $K_a$ modify both the harmonic response and these sideband features. In contrast to the purely $z$-polarized case, $K_c$ also affects the spectra, Fig.~\figref{fig5}{(g)}, because the mixed-polarization drive activates transverse spin motion and qFM contributions. The $K_4$ sweep, Fig.~\figref{fig5}{(h)}, again changes the relative peak heights and asymmetries, showing that the spectra are sensitive to the quartic anisotropy.

Overall, Fig.~\ref{fig5} shows that $z$-polarized spectra provide clear fingerprints of $D$, $K_a$, and $K_4$, but not $K_c$. 
For mixed $y$--$z$-polarized excitation, all three nonlinear magnetization components can become finite. The $z$-component spectra shown in Figs.~\figref{fig5}{(e-h)} already exhibit sensitivity to all four parameters, and the component-resolved benchmarks below further show that the $x$ and $y$ components also carry parameter information. 
This distinction motivates the separate benchmarking of $z$-polarized and mixed-polarization inputs below.
Beyond guiding the choice of input geometry, these parameter sweeps demonstrate a central advantage of nonlinear THz-2DCS over a linear-response analysis based only on the qAFM resonance. In all panels, $J$ is recalibrated to keep the linear qAFM resonance frequency fixed and $\alpha$ is recalibrated to match the target qAFM decay rate. Thus, the spectral changes in Fig.~\ref{fig5} reflect changes in the nonlinear response at fixed linear qAFM resonance frequency and decay rate, rather than trivial shifts or broadening changes of the dominant linear mode.
These features provide additional constraints that make it possible to distinguish parameter sets that would be difficult to separate using the calibrated linear qAFM frequency and decay rate alone.

\begin{figure*}[t!]
\begin{center}
\includegraphics[scale=0.53]{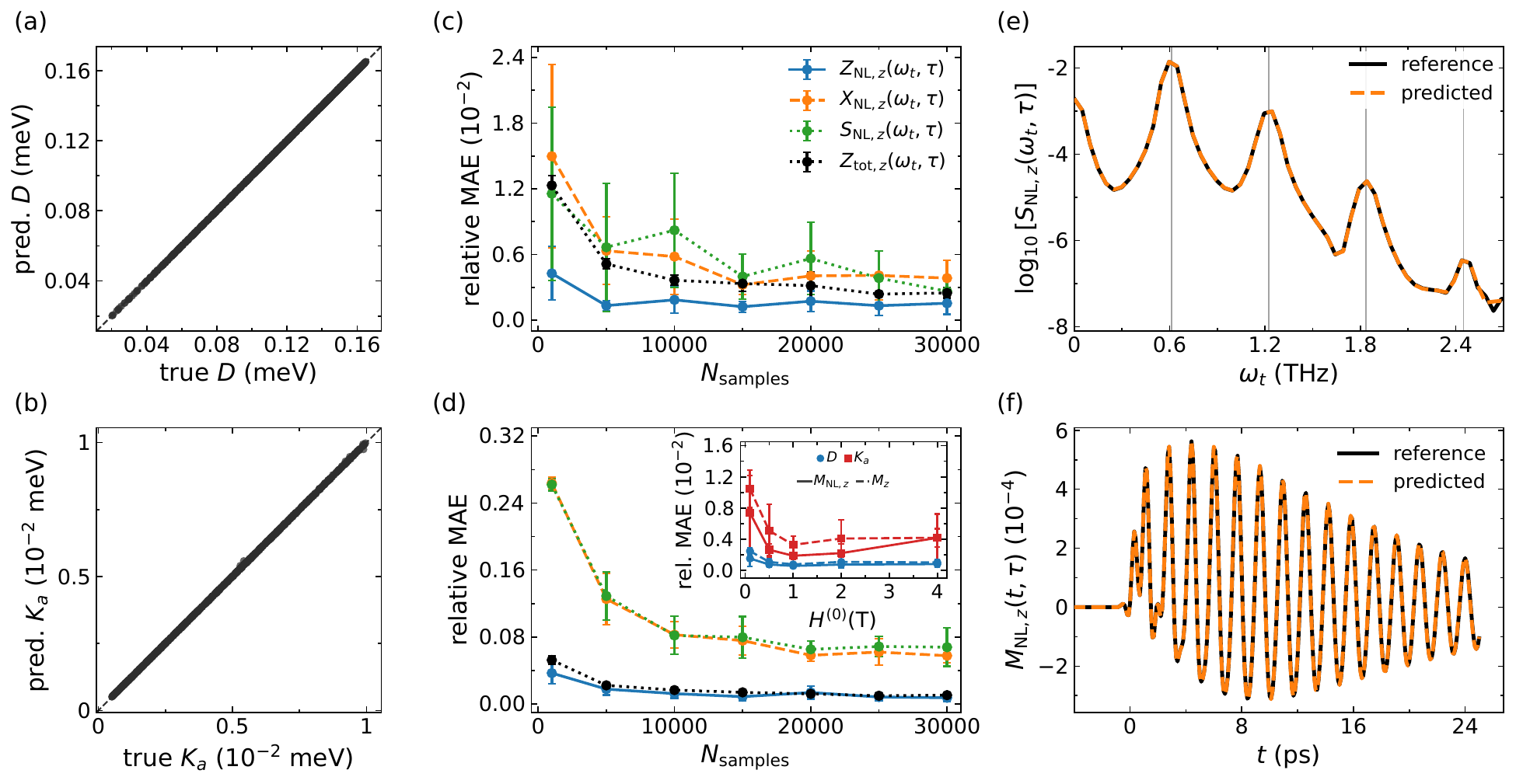}
\caption{
Synthetic-data benchmark for inferring $D$ and $K_a$ from $z$-polarized nonlinear spectra, with $\hat{\mathbf e}_1=\hat{\mathbf e}_2=\hat{\mathbf z}$.
(a,b) True versus predicted values of (a) the DM interaction $D$ and (b) the anisotropy $K_a$ for an independent test set containing $N_{\mathrm{test}}=1000$ simulated spectra, shown for one fixed test-data seed. The network was trained with $N_{\mathrm{samples}}=30000$ spectra at $\tau=4.8$ ps, equal pulse amplitudes $H_1^{(0)}=H_2^{(0)}=1~\mathrm{T}$, and zero added noise. The dashed line indicates perfect prediction. The relative mean absolute errors are $7.3\times10^{-4}$ for $D$ and $2.2\times 10^{-3}$ for $K_a$, while the relative maximum errors are $7.9\times 10^{-3}$ for $D$ and $3.0\times 10^{-2}$ for $K_a$.
(c,d) Relative mean absolute error as a function of training-set size for (c) $D$ and (d) $K_a$. Symbols show the mean over $10$ independently trained networks with different random seeds, and error bars indicate the corresponding standard deviation. Curves compare different choices of input signal and preprocessing. The main input (blue solid curves) is the standardized nonlinear spectrum $Z_{\mathrm{NL},z}(\omega_t,\tau)$ constructed from the nonlinear magnetization $M_{\mathrm{NL},z}(t,\tau)$ using the full preprocessing pipeline. The orange dashed curves use the same nonlinear magnetization input but stop before the final $z$-score standardization, i.e., they use the log-compressed nonlinear spectrum $X_{\mathrm{NL},z}(\omega_t,\tau)$. The green dotted curves use the corresponding Fourier-amplitude nonlinear spectrum $S_{\mathrm{NL},z}(\omega_t,\tau)$ without logarithmic compression or standardization. The black dotted curves use the fully standardized total-magnetization spectrum $Z_{\mathrm{tot},z}(\omega_t,\tau)$ constructed from $M_z(t,\tau)$ rather than from the isolated nonlinear response.
The inset in (d) shows the dependence of the relative error on the THz magnetic-field amplitude $H_{\mathrm{pulse}}$ for $N_{\mathrm{samples}}=30000$, comparing inputs constructed from $M_{\mathrm{NL},z}(t,\tau)$ (solid lines) and $M_z(t,\tau)$ (dashed lines).
(e,f) Forward validation for a selected reference case. The Hamiltonian parameters inferred by the neural network are inserted back into the spin-dynamics forward model and compared with the reference response: (e) nonlinear spectrum and (f) time-domain nonlinear magnetization. For this reference case, the seed-averaged inferred parameters over $10$ independently trained networks are $D_{\mathrm{pred}}=0.1499\pm0.0001~\mathrm{meV}$ and
$K_{a,\mathrm{pred}}=(3.711\pm0.006)\times10^{-3}~\mathrm{meV}$, compared with the reference values $D_{\mathrm{ref}}=0.1499~\mathrm{meV}$ and
$K_{a,\mathrm{ref}}=3.708\times10^{-3}~\mathrm{meV}$.}
\label{fig6}
\end{center}
\end{figure*}


\subsection{Inference from $z$-polarized excitation}
\label{subsec:synthetic_cpol}

We first benchmark Hamiltonian inference for the $z$-polarized excitation geometry, where the nonlinear response is carried by the $z$-axis magnetization component in the reduced two-sublattice model. Motivated by the parameter-sensitivity analysis in Fig.~\ref{fig5}, we begin with a two-parameter inference task for the DM interaction $D$ and the quadratic anisotropy $K_a$. These two parameters show clear spectral fingerprints in $z$-polarized spectra and provide a minimal benchmark for testing whether the CNN can recover Hamiltonian parameters from the standardized nonlinear response.

Figure~\ref{fig6} summarizes the synthetic-data benchmark for the inference task $\bth=(D,K_a)$. Figures~\figref{fig6}{(a,b)} show true-versus-predicted values of $D$ and $K_a$ for an independent test set containing $N_\mathrm{test}=1000$ simulated spectra. The scatter plots are shown for one fixed independently generated test set, corresponding to a single test-data random seed; seed-to-seed variability of the trained networks is quantified separately in the error-bar plots below. The network is trained using $N_{\mathrm{samples}}=30000$ model-development spectra at $\tau=4.8$~ps, equal pulse amplitudes $H_1^{(0)}=H_2^{(0)}=1~\mathrm{T}$, and zero added noise. The points lie close to the diagonal dashed line, demonstrating that the trained CNN accurately predicts both parameters from the nonlinear spectra. Using the error metrics defined in Sec.~\ref{subsec:ml_architecture}, the relative mean absolute errors are $7.3\times10^{-4}$ for $D$ and $2.2\times10^{-3}$ for $K_a$, while the corresponding relative maximum errors are $7.9\times10^{-3}$ and $3.0\times10^{-2}$, respectively. The smaller error for $D$ is consistent with the stronger spectral sensitivity to the DM interaction observed in Fig.~\ref{fig5}.

The dependence on dataset size is shown in Figs.~\figref{fig6}{(c,d)} for $D$ and $K_a$, respectively. We first consider the main input representation used in this work: the standardized log-compressed nonlinear spectrum $Z_{\mathrm{NL},z}(\omega_t,\tau)$ constructed from the isolated nonlinear magnetization $M_{\mathrm{NL},z}(t,\tau)$ (solid blue curves). For this representation, the relative MAE decreases systematically with increasing $N_{\mathrm{samples}}$. Specifically, for $D$, the relative MAE decreases from $4.27\times10^{-3}$ at $N_{\mathrm{samples}}=1000$ to $1.54\times10^{-3}$ at $N_{\mathrm{samples}}=30000$. For $K_a$, the corresponding error decreases from $3.68\times10^{-2}$ to $7.38\times10^{-3}$. The error for $D$ remains smaller than that for $K_a$ over the full range of dataset sizes, again consistent with the stronger spectral sensitivity to $D$ observed in Fig.~\ref{fig5}.

Figures~\figref{fig6}{(c,d)} also compare this main input with modified input representations. The log-compressed nonlinear spectrum $X_{\mathrm{NL},z}(\omega_t,\tau)$ constructed from $M_{\mathrm{NL},z}(t,\tau)$ before the final $z$-score standardization (orange dashed curves) gives noticeably larger errors, especially for $K_a$. This shows that frequency-wise standardization is important for learning the more subtle line-shape and asymmetry changes associated with the anisotropy. 
The nonlinear Fourier-amplitude spectrum $S_{\mathrm{NL},z}(\omega_t,\tau)$ constructed from $M_{\mathrm{NL},z}(t,\tau)$ without logarithmic compression or standardization (green dotted curves) gives comparable errors for $D$ but slightly larger errors for $K_a$ at larger $N_{\mathrm{samples}}$. Thus, using the raw Fourier-amplitude nonlinear spectrum does not uniformly degrade the learning accuracy for all parameters. However, because this representation weights the strongest spectral peaks more heavily, weaker harmonic and line-shape features contribute less directly to the regression. This primarily affects $K_a$, whose spectral fingerprint is more subtle than that of $D$.

The standardized total-magnetization spectrum $Z_{\mathrm{tot},z}(\omega_t,\tau)$ constructed from $M_z(t,\tau)$ rather than from the isolated nonlinear response $M_{\mathrm{NL},z}(t,\tau)$ (black dotted curves) produces slightly larger errors than the nonlinear-input representation. This indicates that the preprocessing pipeline, in particular the logarithmic compression and frequency-wise standardization, can make nonlinear spectral features accessible even when they are embedded in the total magnetization response. Nevertheless, explicitly isolating $M_{\mathrm{NL},z}(t,\tau)$ removes the dominant linear background before preprocessing and therefore makes the relevant nonlinear harmonic and line-shape features more directly available to the network. This distinction becomes especially important in the presence of noise, as discussed in Sec.~\ref{subsec:robustness}: while preprocessing can reveal weak nonlinear features in a total-magnetization spectrum, only the two-pulse nonlinear response provides delay-dependent information that can be exploited to stabilize the inverse problem.

The inset of Fig.~\figref{fig6}{(d)} shows the field-amplitude dependence of the relative errors for inputs constructed from the isolated nonlinear response $M_{\mathrm{NL},z}(t,\tau)$ (solid lines) and from the total magnetization $M_z(t,\tau)$ (dashed lines). For both $D$ and $K_a$, the errors decrease with increasing THz magnetic-field amplitude $H^{(0)}$ up to approximately $1~\mathrm{T}$, reflecting the growth of nonlinear spectral features with drive strength. At larger fields, the errors increase slightly again, likely because strong THz driving modifies the spectral line shape, for example through a redshift of the qAFM resonance, and thereby makes the inverse mapping more complex. Over the full field range, the $M_z(t,\tau)$ input gives larger errors than the corresponding $M_{\mathrm{NL},z}(t,\tau)$ input, especially at low field. This is expected because the total magnetization is dominated by the linear response at weak drive, so the nonlinear spectral features used for inference are less clearly isolated. As $H^{(0)}$ increases, these nonlinear features become more pronounced even in $M_z(t,\tau)$, and the difference between the two input choices decreases.

Figures~\figref{fig6}{(e)} and \figref{fig6}{(f)} provide a forward-validation test. The Hamiltonian parameters inferred by the neural network are inserted back into the spin-dynamics forward model and used to reconstruct the nonlinear response for a selected test example. In both panels, the black solid curve shows the reference response, while the dashed orange curve shows the predicted response obtained from the inferred parameters. The predicted nonlinear spectrum $S_{\mathrm{NL},z}(\omega_t,\tau)$ in Fig.~\figref{fig6}{(e)} and the predicted time-domain nonlinear magnetization $M_{\mathrm{NL},z}(t,\tau)$ in Fig.~\figref{fig6}{(f)} reproduce the reference data with high accuracy. For the test example shown in these panels, the seed-averaged predicted parameters over ten independently trained networks are $D_{\mathrm{pred}}=0.1499\pm0.0001~\mathrm{meV}$ and
$K_{a,\mathrm{pred}}=(3.711\pm0.006)\times10^{-3}~\mathrm{meV}$, while the true parameter values are $D_{\mathrm{ref}}=0.1499~\mathrm{meV}$ and
$K_{a,\mathrm{ref}}=3.708\times10^{-3}~\mathrm{meV}$. The true values lie within the quoted seed-to-seed spread of the predictions, demonstrating accurate parameter inference in addition to the forward reconstruction of the nonlinear THz response.

\begin{figure*}[t!]
\begin{center}
\includegraphics[scale=0.67]{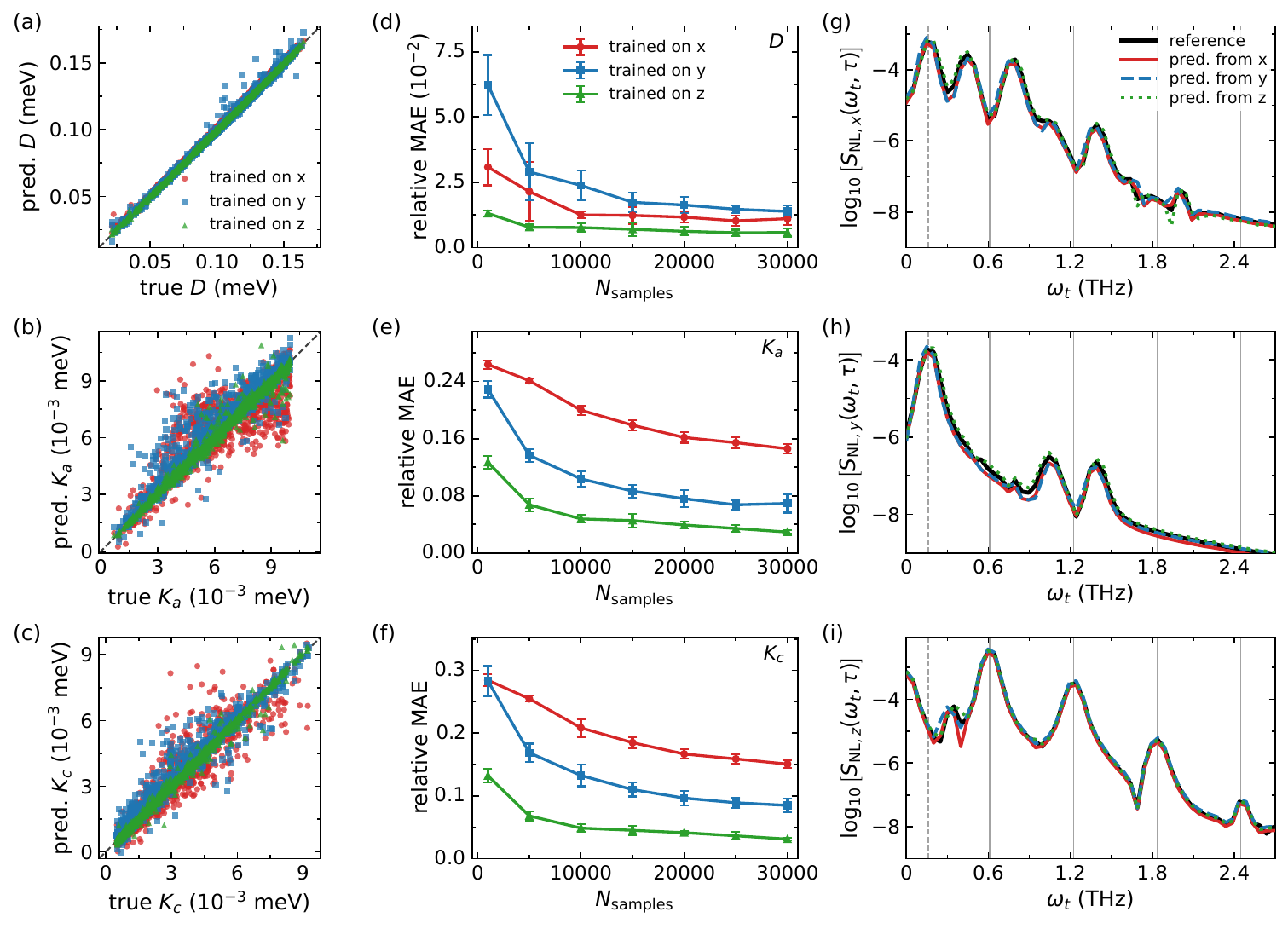}
\caption{
Synthetic-data benchmark for mixed $y$--$z$-polarized excitation.
The networks are trained separately using nonlinear magnetization spectra constructed from the three output components $M_{\mathrm{NL},x}$, $M_{\mathrm{NL},y}$, and $M_{\mathrm{NL},z}$ for equal $y$--$z$-polarized excitation, $\hat{\mathbf e}_1=\hat{\mathbf e}_2=(\hat{\mathbf y}+\hat{\mathbf z})/\sqrt{2}$.
(a--c) True versus predicted values of (a) $D$, (b) $K_a$, and (c) $K_c$ for an independent test set of size $N_{\mathrm{test}}=1000$, shown for one fixed test-data seed. The networks were trained with $N_{\mathrm{samples}}=30000$ simulated spectra at $\tau=4.8$~ps, equal pulse amplitudes $H_1^{(0)}=H_2^{(0)}=1~\mathrm{T}$, and zero added noise. Symbols distinguish models trained on the $x$ (circles), $y$ (squares), or $z$ (triangles) magnetization component. The dashed lines indicate perfect prediction. The relative mean absolute errors for models trained on the $(x,y,z)$ components are
$(1.80\times10^{-2},1.90\times10^{-2},6.27\times10^{-3})$ for $D$,
$(1.68\times10^{-1},1.26\times10^{-1},3.20\times10^{-2})$ for $K_a$, and
$(1.89\times10^{-1},1.67\times10^{-1},4.63\times10^{-2})$ for $K_c$; the corresponding relative maximum errors are
$(6.25\times10^{-1},4.29\times10^{-1},1.86\times10^{-1})$,
$(1.87,2.53,5.20\times10^{-1})$, and $(1.76,2.28,1.03)$, respectively.
(d--f) Relative mean absolute error as a function of training-set size for (d) $D$, (e) $K_a$, and (f) $K_c$. Red, blue, and green curves correspond to networks trained on spectra constructed from $M_{\mathrm{NL},x}(t,\tau)$, $M_{\mathrm{NL},y}(t,\tau)$, and $M_{\mathrm{NL},z}(t,\tau)$, respectively. Symbols show the mean over independently trained networks with ten different random seeds, and error bars indicate the corresponding standard deviation.
(g--i) Forward-validation spectra for a selected test example, showing
(g) $S_{\mathrm{NL},x}(\omega_t,\tau)$, (h) $S_{\mathrm{NL},y}(\omega_t,\tau)$, and (i) $S_{\mathrm{NL},z}(\omega_t,\tau)$. 
In each panel, the black curve shows the reference spectrum, while the solid red, dashed blue, and dotted green curves show predicted spectra obtained from parameters inferred by the models trained on $M_{\mathrm{NL},x}(t,\tau)$, $M_{\mathrm{NL},y}(t,\tau)$, and $M_{\mathrm{NL},z}(t,\tau)$, respectively. 
Solid vertical guide lines mark integer multiples of the qAFM frequency, while the dashed vertical guide line marks the qFM frequency $f_{\mathrm{qFM}}\simeq0.16~\mathrm{THz}$. The mixed-polarization response contains qFM-related nonlinear features: the $z$ component shows spectral weight near $2f_{\mathrm{qFM}}$, while the $x$ and $y$ components show sideband features near $m f_{\mathrm{qAFM}}\pm f_{\mathrm{qFM}}$.
For the test example shown in (g)--(i), the seed-averaged predicted parameters from the
$M_{\mathrm{NL},x}$-, $M_{\mathrm{NL},y}$-, and $M_{\mathrm{NL},z}$-trained models are
$D_{\mathrm{pred}}=(0.154\pm0.001,\,0.157\pm0.001,\,0.156\pm0.001)~\mathrm{meV}$,
$K_{a,\mathrm{pred}}=(1.63\pm0.15,\,1.65\pm0.11,\,1.70\pm0.07)\times10^{-3}~\mathrm{meV}$,
and
$K_{c,\mathrm{pred}}=(0.93\pm0.07,\,0.82\pm0.05,\,0.80\pm0.03)\times10^{-3}~\mathrm{meV}$,
respectively. The true parameter values are
$D_{\mathrm{ref}}=0.156~\mathrm{meV}$,
$K_{a,\mathrm{ref}}=1.71\times10^{-3}~\mathrm{meV}$, and
$K_{c,\mathrm{ref}}=0.84\times10^{-3}~\mathrm{meV}$.
For the $M_{\mathrm{NL},y}$-trained model, all three true values lie within one seed-to-seed standard deviation of the mean prediction; for the $M_{\mathrm{NL},z}$-trained model, this is the case for $D$ and $K_a$, while $K_c$ is slightly outside the one-standard-deviation interval.}
\label{fig7}
\end{center}
\end{figure*}

\subsection{Inference from mixed $y$--$z$-polarized excitation}
\label{subsec:synthetic_bc}

We next benchmark the mixed $y$--$z$-polarized excitation geometry, where
$\hat{\mathbf e}_1=\hat{\mathbf e}_2=(\hat{\mathbf y}+\hat{\mathbf z})/\sqrt{2}$. As discussed in Sec.~\ref{subsec:parameter_sensitivity}, this drive excites both qAFM and qFM dynamics and therefore produces spectral fingerprints of all three parameters considered here, $\bth=(D,K_a,K_c)$. To assess how much information is contained in each nonlinear magnetization component, we train separate neural networks using standardized nonlinear spectra constructed from $M_{\mathrm{NL},x}(t,\tau)$, $M_{\mathrm{NL},y}(t,\tau)$, or $M_{\mathrm{NL},z}(t,\tau)$ as input.

Figures~\figref{fig7}{(a-c)} show true-versus-predicted values of $D$, $K_a$, and $K_c$ for an independent test set containing
$N_{\mathrm{test}}=1000$ spectra. The networks are trained with $N_{\mathrm{samples}}=30000$ spectra at $\tau=4.8~\mathrm{ps}$, equal pulse amplitudes $H_1^{(0)}=H_2^{(0)}=1~\mathrm{T}$, and zero added noise. Symbols distinguish models trained on the
$x$ component, $M_{\mathrm{NL},x}(t,\tau)$ (circles), the $y$ component,
$M_{\mathrm{NL},y}(t,\tau)$ (squares), and the $z$ component,
$M_{\mathrm{NL},z}(t,\tau)$ (triangles). All three Hamiltonian parameters can be inferred from the mixed-polarization spectra, although with component-dependent accuracy. The parameter $D$ is learned more accurately than the anisotropy parameters, consistent with its stronger influence on the nonlinear harmonic response. The successful inference of $K_c$ is a key difference from the purely $z$-polarized case: because the mixed polarization activates transverse qFM dynamics, the $z$-axis anisotropy enters the nonlinear response and becomes learnable.

The accuracy depends strongly on which output component is used as input. Quantitatively, the relative mean absolute errors for the
$M_{\mathrm{NL},x}$-, $M_{\mathrm{NL},y}$-, and $M_{\mathrm{NL},z}$-trained models are
$(1.80\times10^{-2},1.90\times10^{-2},6.27\times10^{-3})$ for $D$,
$(1.68\times10^{-1},1.26\times10^{-1},3.20\times10^{-2})$ for $K_a$, and
$(1.89\times10^{-1},1.67\times10^{-1},4.63\times10^{-2})$ for $K_c$. The corresponding
relative maximum errors are
$(6.25\times10^{-1},4.29\times10^{-1},1.86\times10^{-1})$,
$(1.87,2.53,5.20\times10^{-1})$, and $(1.76,2.28,1.03)$, respectively. Thus, the most accurate results are obtained from the $M_{\mathrm{NL},z}$-trained model, followed by the $M_{\mathrm{NL},y}$- and $M_{\mathrm{NL},x}$-trained models.
This hierarchy is consistent with the mode content of the different response channels. The $z$ component retains a strong qAFM harmonic response while also acquiring qFM-induced sideband features that carry information about $K_c$. The $y$ component contains stronger qFM contributions and sideband features. Although these features are parameter sensitive, different parameter combinations can produce similar qFM frequencies and sideband structures, leading to a less unique inverse mapping. The $x$ component is not directly driven in the equal $y$--$z$ excitation geometry and therefore carries a weaker, more indirect nonlinear fingerprint of the inferred parameters.

The dependence on training-set size, shown in Figs.~\figref{fig7}{(d-f)}, confirms the same hierarchy. Red, blue, and green curves correspond to networks trained on spectra constructed from $M_{\mathrm{NL},x}(t,\tau)$, $M_{\mathrm{NL},y}(t,\tau)$, and $M_{\mathrm{NL},z}(t,\tau)$, respectively. For all three inferred parameters, the relative MAE decreases as $N_{\mathrm{samples}}$ increases. The $M_{\mathrm{NL},z}$ input yields the lowest errors over the full range of training-set sizes, whereas $M_{\mathrm{NL},y}$ and $M_{\mathrm{NL},x}$ produce larger errors. Among the latter two, $M_{\mathrm{NL},y}$ performs better in learning $K_a$ and $K_c$. This confirms that the $z$-component nonlinear magnetization retains the most informative combination of qAFM harmonic features and qFM-induced sideband structure for this excitation geometry. 

Figures~\figref{fig7}{(g-i)} show the forward-model comparison for a selected test example across all three nonlinear output components. The panels compare the reference nonlinear spectra with predicted spectra obtained by inserting the inferred parameters back into the spin-dynamics forward model. Specifically, they show (g) $S_{\mathrm{NL},x}(\omega_t,\tau)$, (h) $S_{\mathrm{NL},y}(\omega_t,\tau)$, and (i) $S_{\mathrm{NL},z}(\omega_t,\tau)$. In each panel, the black curve is the reference spectrum. The vertical guide lines mark integer multiples of the qAFM frequency (solid lines) and the qFM frequency (dashed line), illustrating that the mixed-polarization response contains both qAFM harmonic structure and qFM-related sideband features. In particular, the $x$ and $y$ components contain sidebands near $m f_{\mathrm{qAFM}}\pm f_{\mathrm{qFM}}$, while the $z$ component contains spectral weight near $2f_{\mathrm{qFM}}$. The solid red curves use parameters inferred by the $M_{\mathrm{NL},x}$-trained model, the dashed blue curves use parameters inferred by the $M_{\mathrm{NL},y}$-trained model, and the dotted green curves use parameters inferred by the $M_{\mathrm{NL},z}$-trained model. The predicted spectra closely reproduce the reference response for all three output components, showing that the inferred parameter sets generate nonlinear dynamics consistent with the reference spectra, even though the individual parameters are not learned with identical accuracy.

For the test example shown in Figs.~\figref{fig7}{(g-i)}, the seed-averaged predicted parameters from the
$M_{\mathrm{NL},x}$-, $M_{\mathrm{NL},y}$-, and $M_{\mathrm{NL},z}$-trained models are
$D_{\mathrm{pred}}=(1.54\pm0.01,\,1.57\pm0.01,\,1.56\pm0.01)\times10^{-1}~\mathrm{meV}$,
$K_{a,\mathrm{pred}}=(1.63\pm0.15,\,1.65\pm0.11,\,1.70\pm0.07)\times10^{-3}~\mathrm{meV}$, and
$K_{c,\mathrm{pred}}=(0.93\pm0.07,\,0.82\pm0.05,\,0.80\pm0.03)\times10^{-3}~\mathrm{meV}$, respectively. The true parameter values are
$D_{\mathrm{ref}}=1.56\times10^{-1}~\mathrm{meV}$,
$K_{a,\mathrm{ref}}=1.71\times10^{-3}~\mathrm{meV}$, and
$K_{c,\mathrm{ref}}=0.84\times10^{-3}~\mathrm{meV}$.
The quoted uncertainties denote the seed-to-seed standard deviation over the ten independently trained networks. With this criterion, all three true values lie within one standard deviation of the mean prediction for the $M_{\mathrm{NL},y}$-trained model. For the $M_{\mathrm{NL},z}$-trained model, this is the case for $D$ and $K_a$, while $K_c$ is slightly outside the one-standard-deviation interval; for the $M_{\mathrm{NL},x}$-trained model, only $K_a$ lies within one standard deviation. 
Overall, the forward-model comparison shows that parameter sets inferred from component-resolved nonlinear spectra can reproduce the selected test response. However, the full test-set statistics show that the $M_{\mathrm{NL},z}$ input provides the most robust global inference accuracy. The reduced accuracy for the $M_{\mathrm{NL},x}$ and $M_{\mathrm{NL},y}$ inputs is consistent with the weaker or less unique fingerprints discussed above: the $x$ response is generated indirectly, while qFM-related features in the $y$ response can be produced by different parameter combinations with similar qFM frequencies or sideband structures.

\begin{figure*}[t!]
\begin{center}
\includegraphics[scale=0.51]{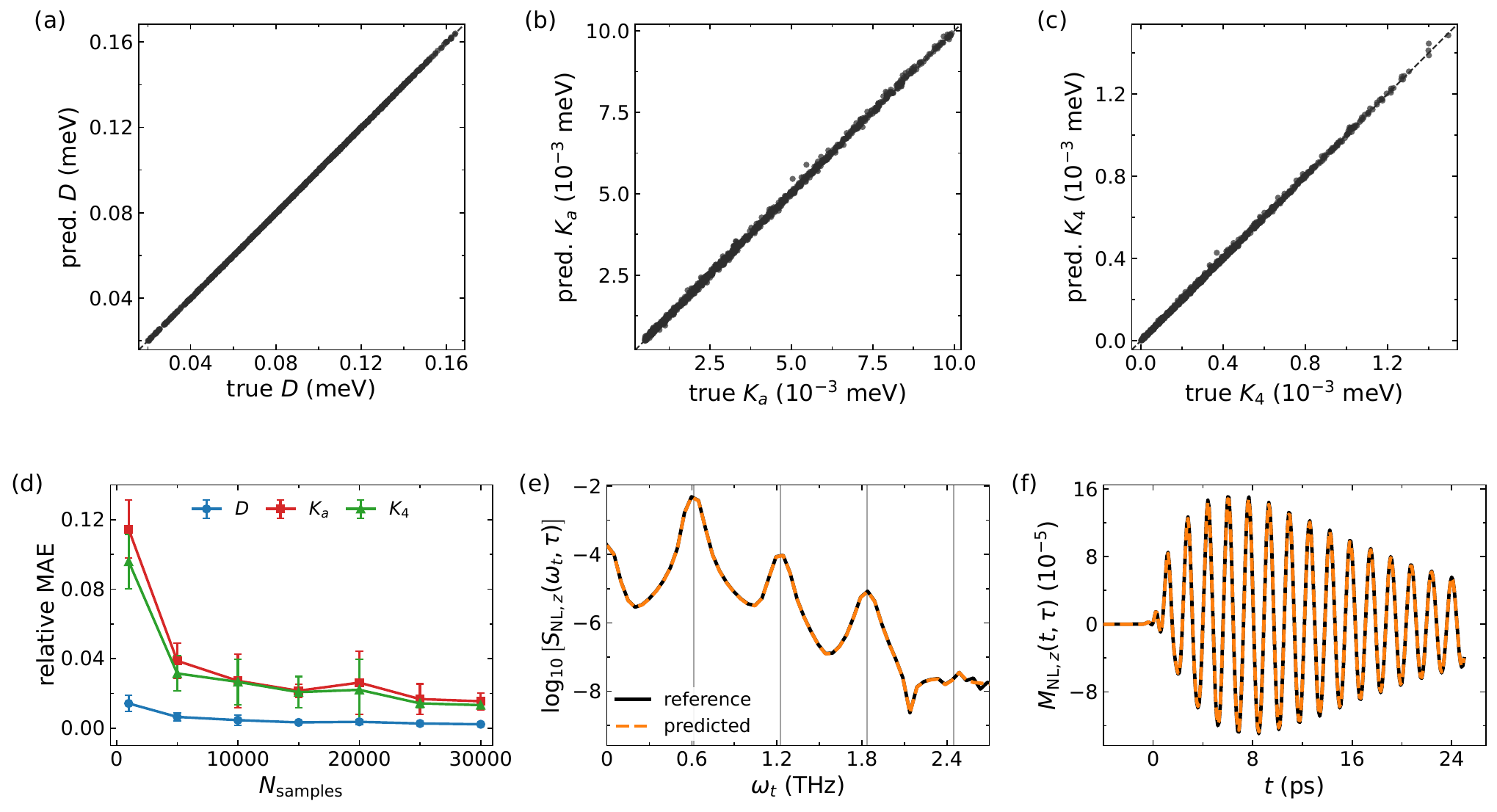}
\caption{
Synthetic-data benchmark for inferring $D$, $K_a$, and $K_4$ from $z$-polarized nonlinear spectra, with $\hat{\mathbf e}_1=\hat{\mathbf e}_2=\hat{\mathbf z}$.
(a--c) True versus predicted values of (a) the DM interaction $D$, (b) the quadratic anisotropy $K_a$, and (c) the quartic anisotropy $K_4$ for an independent test set of size $N_{\mathrm{test}}=1000$, shown for one fixed test-data seed. The network was trained using the standardized nonlinear spectrum $Z_{\mathrm{NL},z}(\omega_t,\tau)$ constructed from $M_{\mathrm{NL},z}(t,\tau)$, with $N_{\mathrm{samples}}=30000$ simulated spectra at $\tau=4.8~\mathrm{ps}$, equal pulse amplitudes $H_1^{(0)}=H_2^{(0)}=0.5~\mathrm{T}$, and zero added noise. The dashed line indicates perfect prediction. The relative mean absolute errors are
$2.77\times10^{-3}$ for $D$, $1.87\times10^{-2}$ for $K_a$, and $2.46\times10^{-2}$ for $K_4$, while the corresponding relative maximum errors are $2.79\times10^{-2}$, $2.14\times10^{-1}$, and $7.30\times10^{-1}$, respectively.
(d) Relative mean absolute error as a function of training-set size for $D$ (blue), $K_a$ (red), and $K_4$ (green). Symbols show the mean over ten independently trained networks with different random seeds, and error bars indicate the corresponding standard deviation.
(e,f) Forward validation for a selected test example. The Hamiltonian parameters inferred by the neural network are inserted back into the spin-dynamics forward model, and the predicted response is compared with the reference data. Panel (e) shows the reference nonlinear spectrum $S_{\mathrm{NL},z}(\omega_t,\tau)$ (solid black line) and the predicted nonlinear spectrum (dashed orange line), while panel (f) shows the corresponding time-domain nonlinear magnetization $M_{\mathrm{NL},z}(t,\tau)$.
For the test example shown in (e,f), the seed-averaged predicted parameters over $10$ independently trained networks are
$D_{\mathrm{pred}}=(6.82\pm0.03)\times10^{-2}~\mathrm{meV}$,
$K_{a,\mathrm{pred}}=(7.2\pm0.4)\times10^{-4}~\mathrm{meV}$, and
$K_{4,\mathrm{pred}}=(9.7\pm0.1)\times10^{-4}~\mathrm{meV}$.
The corresponding true parameter values are
$D_{\mathrm{ref}}=6.81\times10^{-2}~\mathrm{meV}$,
$K_{a,\mathrm{ref}}=6.8\times10^{-4}~\mathrm{meV}$, and
$K_{4,\mathrm{ref}}=9.7\times10^{-4}~\mathrm{meV}$.
For all three parameters, the true values lie within one seed-to-seed standard deviation of the mean prediction.
}
\label{fig8}
\end{center}
\end{figure*}

\subsection{Inference of quartic anisotropy from $z$-polarized excitation}
\label{subsec:synthetic_k4}

We next extend the $z$-polarized benchmark to a three-parameter inference task, $\bth=(D,K_a,K_4)$. This tests whether the nonlinear spectra contain enough information not only to distinguish the DM interaction and the quadratic anisotropy, but also to infer the quartic anisotropy that controls higher-order anisotropy effects in the nonlinear spin dynamics. The input is again the standardized nonlinear spectrum $Z_{\mathrm{NL},z}(\omega_t,\tau)$ constructed from the isolated nonlinear magnetization $M_{\mathrm{NL},z}(t,\tau)$. The results in Fig.~\ref{fig8} are obtained with $N_{\mathrm{samples}}=30000$ simulated spectra at $\tau=4.8~\mathrm{ps}$, equal pulse amplitudes $H_1^{(0)}=H_2^{(0)}=0.5~\mathrm{T}$, $z$-polarized excitation with $\hat{\mathbf e}_1=\hat{\mathbf e}_2=\hat{\mathbf z}$, and zero added noise.

Figures~\figref{fig8}{(a-c)} show true-versus-predicted values of $D$, $K_a$, and $K_4$ for an independent test set of size $N_\mathrm{test}=1000$. The DM interaction remains the most accurately learned parameter, with relative MAE $2.77\times10^{-3}$ and relative maximum error $2.79\times10^{-2}$. The quadratic anisotropy $K_a$ is learned with relative MAE $1.87\times10^{-2}$ and relative maximum error $2.14\times10^{-1}$. The quartic anisotropy $K_4$ is also captured, with relative MAE $2.46\times10^{-2}$, although its relative maximum error is larger, at $7.30\times10^{-1}$. The larger maximum relative error for $K_4$ arises from an isolated test example at very small $K_4$, where a modest absolute deviation is amplified when expressed as a relative error.

The dependence on training-set size, shown in Fig.~\figref{fig8}{(d)}, exhibits the same trend. The blue, red, and green curves show the relative MAE for $D$ (blue), $K_a$ (red), and $K_4$ (green), respectively. For all three parameters, the relative MAE decreases as $N_{\mathrm{samples}}$ is increased, showing that the inference improves systematically with training-library size. The $D$ error remains the smallest over the full range of training-set sizes, while $K_a$ and $K_4$ show larger comparable errors. This behavior is consistent with the fact that $D$ has a strong and direct influence on the nonlinear harmonic response, whereas $K_a$ and $K_4$ mainly affect more subtle line-shape changes, peak asymmetries, and relative harmonic intensities, as observable in Fig.~\ref{fig5}.

Figures~\figref{fig8}{(e-f)} present the forward-validation check for the $z$-polarized inference task with $\bth=(D,K_a,K_4)$. As above, the Hamiltonian parameters inferred by the neural network are inserted back into the spin-dynamics forward model and used to predict the nonlinear response for a selected test example. In Fig.~\figref{fig8}{(e)}, the reference nonlinear spectrum $S_{\mathrm{NL},z}(\omega_t,\tau)$ (solid black line) is compared with the predicted spectrum obtained from the inferred parameters (dashed orange line). The agreement demonstrates that the inferred parameter set reproduces the main spectral fingerprint of the reference dynamics. Figure~\figref{fig8}{(f)} shows the corresponding time-domain nonlinear magnetization $M_{\mathrm{NL},z}(t,\tau)$, with the same reference and prediction comparison. The predicted time trace also follows the reference response closely, confirming that the inferred parameters reproduce both the spectral and time-domain nonlinear dynamics for this representative example.

For the test example shown in Figs.~\figref{fig8}{(e-f)}, the seed-averaged predicted parameters over ten independently trained networks are
$D_{\mathrm{pred}}=(6.82\pm0.03)\times10^{-2}~\mathrm{meV}$,
$K_{a,\mathrm{pred}}=(7.2\pm0.4)\times10^{-4}~\mathrm{meV}$, and
$K_{4,\mathrm{pred}}=(9.7\pm0.1)\times10^{-4}~\mathrm{meV}$. The corresponding true parameter values are
$D_{\mathrm{ref}}=6.81\times10^{-2}~\mathrm{meV}$,
$K_{a,\mathrm{ref}}=6.8\times10^{-4}~\mathrm{meV}$, and
$K_{4,\mathrm{ref}}=9.7\times10^{-4}~\mathrm{meV}$. For all three parameters, the true value lies within one seed-to-seed standard deviation of the mean prediction. Thus, the $z$-polarized nonlinear spectrum supports inference of the quartic anisotropy in addition to the lower-order Hamiltonian parameters, although $K_4$ remains more challenging to learn than $D$.

\begin{figure*}[t!]
\begin{center}
\includegraphics[scale=0.45]{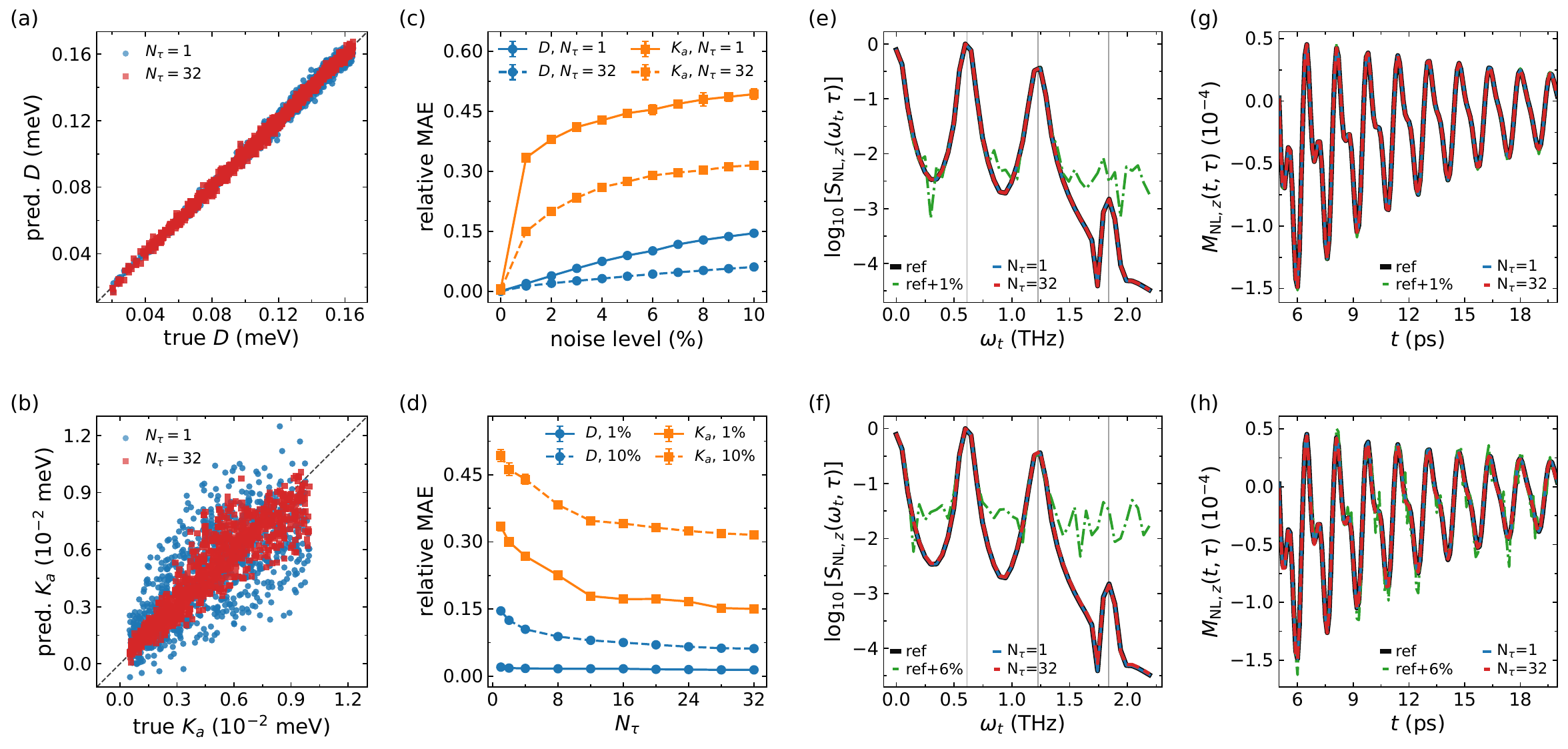}
\caption{
Effect of time-domain noise and number of included delays.
All panels use $z$-polarized excitation with
$\hat{\mathbf e}_1=\hat{\mathbf e}_2=\hat{\mathbf z}$ and equal pulse amplitudes
$H_1^{(0)}=H_2^{(0)}=1~\mathrm{T}$.
The neural networks are trained using $N_{\mathrm{samples}}=30000$ simulated spectra constructed from the $z$-component nonlinear magnetization. Gaussian noise is added in the time domain before applying the Hann window and Fourier transform; $N_\tau=1$ denotes a single-delay input, while $N_\tau=32$ denotes the full multi-delay input.
(a,b) True versus predicted values of (a) $D$ and (b) $K_a$ for an independent noisy test set with $1\%$ time-domain noise, shown for one fixed noisy test-data seed. Blue circles show predictions using $N_\tau=1$, while red squares show predictions using $N_\tau=32$. The dashed line indicates perfect prediction. The relative mean absolute errors, ordered as $(N_\tau=1,N_\tau=32)$, are $(2.10\times10^{-2},1.80\times10^{-2})$ for $D$ and $(4.12\times10^{-1},1.65\times10^{-1})$ for $K_a$, while the corresponding relative maximum errors are $(1.68\times10^{-1},1.96\times10^{-1})$ and $(3.91,1.07)$, respectively.
(c) Relative mean absolute error as a function of noise level for $D$ (blue) and $K_a$ (orange), comparing single-delay inputs, $N_\tau=1$ (solid lines), and full multi-delay inputs, $N_\tau=32$ (dashed lines). Symbols show the mean over independently trained networks with ten different random seeds, and error bars indicate the corresponding standard deviation. At $10\%$ noise, the relative MAE values for $(D,K_a)$ are $(1.46\times10^{-1},4.93\times10^{-1})$ for $N_\tau=1$ and $(6.15\times10^{-2},3.16\times10^{-1})$ for $N_\tau=32$.
(d) Relative mean absolute error as a function of the number of included delays for $1\%$ noise (solid lines) and $10\%$ noise (dashed lines), shown for $D$ (blue) and $K_a$ (orange). Increasing $N_\tau$ reduces the prediction error, demonstrating that the delay dependence of the nonlinear response provides additional information for more robust Hamiltonian inference.
(e,f) Forward-validation spectra for selected test examples with (e) $1\%$ and (f) $6\%$ noise. The nonlinear spectra $S_{\mathrm{NL},z}(\omega_t,\tau)$ are shown at $\tau=2.4~\mathrm{ps}$, normalized to the qAFM peak near $0.61~\mathrm{THz}$, and plotted on a logarithmic scale. The clean reference spectrum (black solid line) is compared with the noisy reference spectrum (green dash--dotted line), and with predicted spectra obtained from parameters inferred using $N_\tau=1$ (blue solid line) and $N_\tau=32$ (red dashed line).
(g,h) Corresponding time-domain nonlinear magnetization dynamics $M_{\mathrm{NL},z}(t,\tau)$ for the same cases as in (e,f), shown in the after-pulse window with the same line convention. For the $1\%$-noise case in (e,g), the seed-averaged predicted parameters are $D_{\mathrm{pred}}=(1.41\pm0.03,\,1.44\pm0.02)\times10^{-1}~\mathrm{meV}$ and $K_{a,\mathrm{pred}}=(4.4\pm1.8,\,2.8\pm0.3)\times10^{-3}~\mathrm{meV}$ for $(N_\tau=1,N_\tau=32)$, respectively. For the $6\%$-noise case in (f,h), the corresponding values are $D_{\mathrm{pred}}=(1.39\pm0.14,\,1.48\pm0.08)\times10^{-1}~\mathrm{meV}$ and $K_{a,\mathrm{pred}}=(3.4\pm2.4,\,2.9\pm0.9)\times10^{-3}~\mathrm{meV}$. The true parameter values are $D_{\mathrm{ref}}=1.45\times10^{-1}~\mathrm{meV}$ and $K_{a,\mathrm{ref}}=2.8\times10^{-3}~\mathrm{meV}$. For the multi-delay input, the true values lie within one seed-to-seed standard deviation of the mean prediction for both noise levels.
}
\label{fig9}
\end{center}
\end{figure*}

\subsection{Robustness to time-domain noise and delay sampling}
\label{subsec:robustness}

To assess the robustness of the inverse model under experimentally more realistic conditions, we add noise directly to the time-domain nonlinear response before constructing the spectral input. The noise is added after subtracting the pre-pulse baseline and after selecting the detection window, but before applying the Hann window and Fourier transform. For a learning input constructed from the magnetization component $\nu$ and a set of included delays $\mathcal T$, the noise scale is defined from the clean, baseline-corrected signal of the same component used for learning,
\begin{align}
\sigma_{\mathrm{noise}}^{(\nu,\mathcal T)}
=
\frac{p}{100}
\max_{\tau\in\mathcal T}
\max_{m}
\left|
\widetilde M_{\mathrm{NL},\nu}(t_m,\tau)
\right|\,,
\label{eq:noise_sigma}
\end{align}
where $p$ is the noise level in percent and the maximum over $m$ is taken within the selected after-pulse detection window. Thus, for a single-delay input, $\mathcal T$ contains only one delay, whereas for a multi-delay input the same noise scale is used for all delays included in the input. In the robustness benchmarks below, the input is constructed from the $z$ component, so Eq.~\eqref{eq:noise_sigma} is evaluated using $\widetilde M_{\mathrm{NL},z}(t_m,\tau)$.

The noisy time-domain signal is then
\begin{align}
M_{\mathrm{NL},\nu}^{\mathrm{noise}}(t_m,\tau)
=
\widetilde M_{\mathrm{NL},\nu}(t_m,\tau)
+
\eta_\nu(t_m,\tau)\,,
\end{align}
where $\eta_\nu(t_m,\tau)$ is independent Gaussian noise with zero mean and standard deviation $\sigma_{\mathrm{noise}}^{(\nu,\mathcal T)}$. The noisy trace is multiplied by the same Hann window used in the clean preprocessing pipeline,
\begin{align}
M_{\mathrm{NL},\nu}^{\mathrm{noise,win}}(t_m,\tau)
=
w_m
M_{\mathrm{NL},\nu}^{\mathrm{noise}}(t_m,\tau)\,,
\end{align}
after which the Fourier transform, normalization, logarithmic compression, and training-set $z$-score standardization are applied as before. This procedure treats the noise as time-domain measurement noise in the after-pulse response, rather than as an artificial perturbation added directly to the already processed spectrum. Examples of the resulting noisy time-domain traces are shown in Figs.~\figref{fig9}{(g-h)}, where the clean reference dynamics are shown by the black solid curves and the corresponding noisy traces by the green dash--dotted curves.

Figure~\ref{fig9} summarizes the effect of time-domain noise and the number of included delays on the two-parameter inference task $\bth=(D,K_a)$. We compare a single-delay input, $N_\tau=1$, at $\tau=2.4~\mathrm{ps}$ with the full multi-delay input, $N_\tau=32$, which contains equally spaced delays from $\tau=2.4$ to $5.5~\mathrm{ps}$ with spacing $\Delta\tau=0.1~\mathrm{ps}$.  Panels~\figref{fig9}{(a-b)} show true-versus-predicted values of $D$ and $K_a$ for an independent test set with $1\%$ added time-domain noise, shown for representative trained networks. Blue circles correspond to the single-delay input and red squares to the full multi-delay input. The scatter is reduced most clearly for $K_a$ when the full delay set is used. For the trained networks shown in these panels, the relative mean absolute errors, ordered as $(N_\tau=1,N_\tau=32)$, are $(2.10\times10^{-2},1.80\times10^{-2})$ for $D$ and $(4.12\times10^{-1},1.65\times10^{-1})$ for $K_a$. The corresponding relative maximum errors are $(1.68\times10^{-1},1.96\times10^{-1})$ and $(3.91,1.07)$, respectively. Thus, adding delay-dependent information gives a modest improvement in the mean error for $D$ at this noise level, but a much stronger improvement for $K_a$. This difference is consistent with $K_a$ primarily manifesting through subtle line-shape distortions and spectral asymmetries, rather than through prominent harmonic features.

The systematic noise dependence is shown in Fig.~\figref{fig9}{(c)}. The blue and orange curves correspond to the relative MAE for $D$ and $K_a$, respectively, while the solid and dashed lines distinguish the single-delay case ($N_\tau=1$) from the full multi-delay input ($N_\tau=32$). The symbols show the mean over ten independently trained networks, and the error bars indicate the corresponding seed-to-seed standard deviation. For both parameters, the relative MAE increases with noise level, as expected. However, the increase is markedly suppressed for the full multi-delay input compared with the single-delay input. At $10\%$ noise, the relative MAE values, ordered as $(D,K_a)$, are $(1.46\times10^{-1},4.93\times10^{-1})$ for $N_\tau=1$, but decrease to $(6.15\times10^{-2},3.16\times10^{-1})$ for $N_\tau=32$. This shows that the network does not rely only on one noisy spectrum at a single delay, but can exploit the correlated delay dependence of the nonlinear response.

Figure~\figref{fig9}{(d)} presents the same trend from the complementary perspective: the relative MAE is plotted as a function of the number of included delays $N_\tau$. Solid lines show the $1\%$-noise case and dashed lines show the $10\%$-noise case, with blue and orange markers again corresponding to $D$ and $K_a$. Increasing $N_\tau$ reduces the prediction error for both parameters. For $1\%$ noise, the relative MAE values, ordered as $(D,K_a)$, decrease from
$(2.02\times10^{-2},3.34\times10^{-1})$ at $N_\tau=1$ to
$(1.40\times10^{-2},1.50\times10^{-1})$ at $N_\tau=32$. For $10\%$ noise, the corresponding decrease is from $(1.46\times10^{-1},4.93\times10^{-1})$ to
$(6.15\times10^{-2},3.16\times10^{-1})$. The improvement is particularly important for $K_a$, whose spectral fingerprint is more subtle than that of $D$ and is therefore more sensitive to time-domain noise.

Figures~\figref{fig9}{(e-h)} provide forward-validation examples for selected noisy test examples at $\tau=2.4~\mathrm{ps}$. Figures~\figref{fig9}{(e,f)} compare nonlinear spectra $S_{\mathrm{NL},z}(\omega_t,\tau)$ for $1\%$ and $6\%$ noise, respectively, while Figs.~\figref{fig9}{(g,h)} show the corresponding time-domain nonlinear magnetization dynamics $M_{\mathrm{NL},z}(t,\tau)$. The clean reference response is shown by the black solid line and the noisy reference response by the green dash--dotted line. The blue solid and red dashed curves show predicted responses obtained from parameters inferred using $N_\tau=1$ and $N_\tau=32$, respectively. 
For both noise levels, the predicted spectra and time-domain dynamics agree well with the clean reference response, even though the inputs used for inference are constructed from noisy time-domain traces. This shows that the inferred parameters remain physically meaningful: when inserted back into the forward model, they reproduce the underlying clean nonlinear response rather than only fitting noise-distorted input spectra.

For the $1\%$-noise case in Figs.~\figref{fig9}{(e,g)}, the seed-averaged predicted parameters are
$D_{\mathrm{pred}}=(1.41\pm0.03,\,1.44\pm0.02)\times10^{-1}~\mathrm{meV}$ and
$K_{a,\mathrm{pred}}=(4.4\pm1.8,\,2.8\pm0.3)\times10^{-3}~\mathrm{meV}$ for
$(N_\tau=1,N_\tau=32)$, respectively. For the $6\%$-noise case in
Figs.~\figref{fig9}{(f,h)}, the corresponding values are
$D_{\mathrm{pred}}=(1.39\pm0.14,\,1.48\pm0.08)\times10^{-1}~\mathrm{meV}$ and
$K_{a,\mathrm{pred}}=(3.4\pm2.4,\,2.9\pm0.9)\times10^{-3}~\mathrm{meV}$. The true parameter values are $D_{\mathrm{ref}}=1.45\times10^{-1}~\mathrm{meV}$ and $K_{a,\mathrm{ref}}=2.8\times10^{-3}~\mathrm{meV}$. For the multi-delay input, the seed-averaged predictions are consistent with the true values within the quoted seed-to-seed standard deviation for both noise levels. This supports the conclusion from Figs.~\figref{fig9}{(c,d)} that including the delay dependence improves the robustness of the inference.

\begin{figure}[t!]
\begin{center}
\includegraphics[scale=0.52]{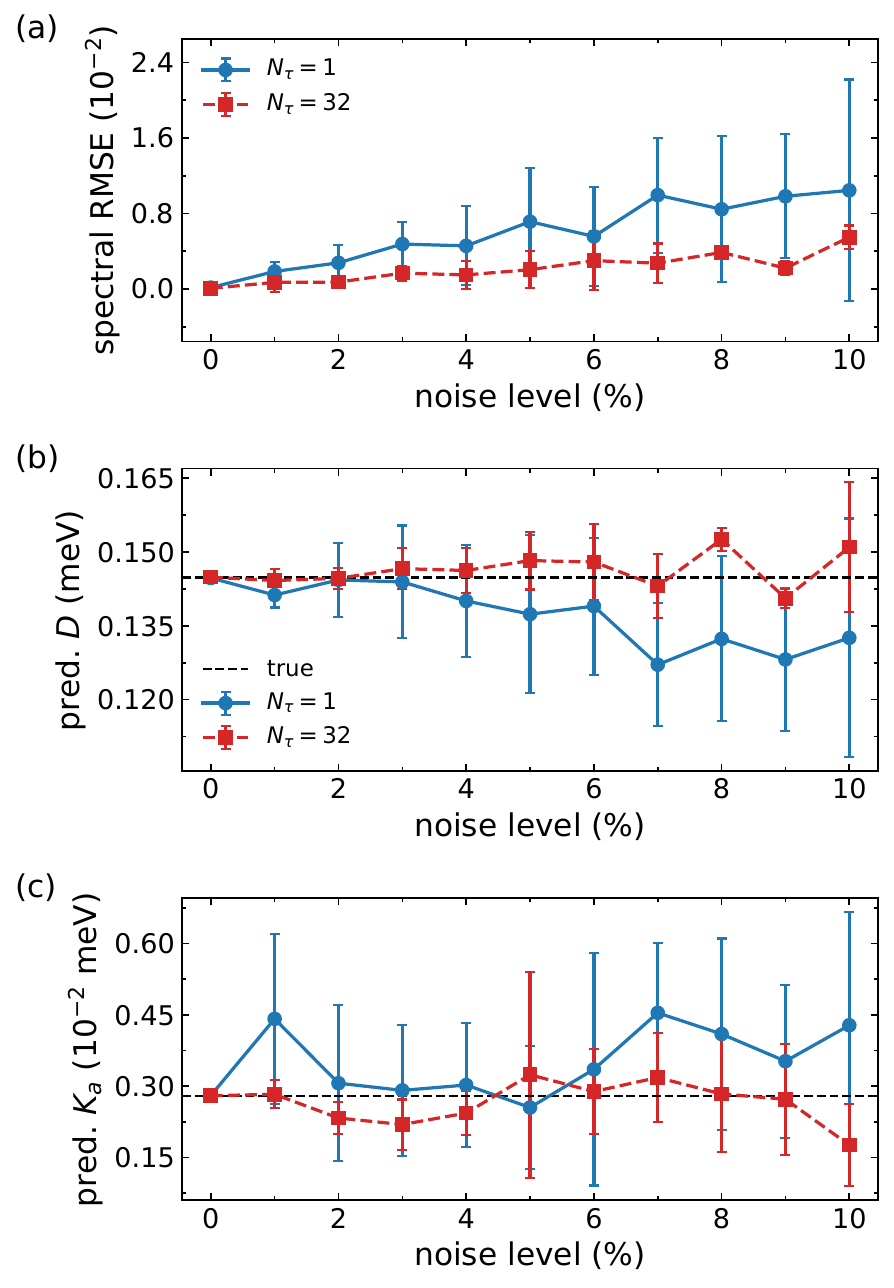}
\caption{
Noise dependence of forward-validation accuracy and inferred parameters.
(a) Spectral prediction error, quantified as the root-mean-square error (RMSE), as a function of the time-domain noise level for single-delay input, $N_\tau=1$ (blue), and full multi-delay input, $N_\tau=32$ (red). The RMSE is computed between the clean reference nonlinear spectrum $S_{\mathrm{NL},z}(\omega_t,\tau)$ and the predicted spectrum obtained by inserting the inferred parameters back into the forward model. Symbols show the mean over independently trained networks with ten different random seeds, and error bars indicate the corresponding seed-to-seed standard deviation.
(b,c) Seed-averaged inferred values of (b) $D$ and (c) $K_a$ as a function of noise level for $N_\tau=1$ (blue) and $N_\tau=32$ (red). Error bars indicate the seed-to-seed standard deviation, and the dashed horizontal line indicates the true value of the corresponding Hamiltonian parameter. The comparison shows that using the full multi-delay input stabilizes the predicted spectra and reduces the noise-induced deviations of the inferred parameters relative to the single-delay input.
}
\label{fig10}
\end{center}
\end{figure}

Figure~\ref{fig10} further quantifies how time-domain noise affects the forward-validation accuracy and the inferred parameters. Figure~\figref{fig10}{(a)} shows the spectral prediction error, quantified as the root-mean-square error (RMSE) between the clean reference nonlinear spectrum $S_{\mathrm{NL},z}(\omega_t,\tau)$ and the predicted spectrum obtained from the inferred parameters. The RMSE increases with noise for both input choices, but the full multi-delay input, $N_\tau=32$ (red), consistently gives a smaller spectral prediction error than the single-delay input, $N_\tau=1$ (blue). In addition, the error bars are substantially smaller for the multi-delay input, indicating reduced seed-to-seed variability of the forward prediction. Thus, the improvement provided by the delay-dependent input is not limited to the direct parameter-error metrics; it also leads to a more accurate and more stable forward prediction of the nonlinear spectrum.

Figures~\figref{fig10}{(b,c)} show the corresponding seed-averaged inferred values of $D$ and $K_a$ as a function of noise level. The dashed horizontal lines mark the true parameter values. For both parameters, the multi-delay input reduces the noise-induced deviation of the seed-averaged predictions relative to the single-delay input. The error bars also show that the multi-delay input generally stabilizes the inference against seed-to-seed variations, especially at elevated noise levels. These results confirm that delay-dependent information improves not only the visual agreement of the forward spectra but also the stability of the inferred Hamiltonian parameters.

In summary, Figs.~\ref{fig9} and \ref{fig10} demonstrate that the inverse model remains robust under time-domain noise, but that the robustness depends strongly on the number of delays supplied to the network. A single nonlinear spectrum already constrains the dominant parameter $D$, whereas the full multi-delay THz-2DCS input substantially improves the inference of the more noise-sensitive anisotropy $K_a$ and also improves the robustness of $D$ at higher noise levels. The improvement obtained by increasing $N_\tau$ demonstrates the value of the second spectroscopic dimension: the network can use the correlated evolution of nonlinear spectral features with inter-pulse delay, rather than relying on one spectrum alone. This delay-dependent information provides additional experimentally accessible constraints and stabilizes the inverse problem against noise.

\section{Application to experimental THz-2DCS data}
\label{sec:experiment_application}

We now apply the Hamiltonian-inference framework developed above to experimental THz data from Sm$_{0.4}$Er$_{0.6}$FeO$_3$~\cite{Huang2024}. The measurements were performed on the same $b$-cut single crystal previously characterized by THz time-domain spectroscopy~\cite{Zhao2016}. At room temperature, the material is above the spin-reorientation region and is therefore in the $\Gamma_4$ configuration, with the antiferromagnetic vector predominantly along the crystallographic $a$ axis and the weak ferromagnetic moment along the $c$ axis. Using the coordinate convention introduced above, this corresponds to dominant antiferromagnetic order along $x$ and weak ferromagnetism along $z$. In this configuration, a THz magnetic field polarized along $z$ couples strongly to the qAFM resonance, which appears near $0.61~\mathrm{THz}$ in the transmitted THz response~\cite{Huang2024}. This makes the room-temperature data a natural experimental test case for the $z$-polarized inference protocol developed above.

We apply two trained inverse models to the experimental nonlinear THz response. The first model infers the two-parameter set $\bth=(D,K_a)$, while the second model infers the three-parameter set $\bth=(D,K_a,K_4)$. Both models use the same excitation geometry as the experiment, namely $z$-polarized THz magnetic fields with $\hat{\mathbf e}_1=\hat{\mathbf e}_2=\hat{\mathbf z}$, and take as input the standardized nonlinear $z$-component spectra, $Z_{\mathrm{NL},z}(\omega_t,\tau)$, for the three-delay set $\mathcal T=\{4.0,4.4,4.8\}~\mathrm{ps}$. Comparing the two models allows us to test whether including the quartic anisotropy improves the reconstruction of the experimental nonlinear response.

Before applying the neural network, we first determine the experimental inputs $\boldsymbol{\eta}$ that define the forward model. These include the temporal profiles of the two THz excitation pulses and the linear qAFM response used to calibrate the exchange scale $J$ and the Gilbert damping parameter $\alpha$ in the simulations. The two experimentally measured pulse waveforms are fitted to chirped Gaussian-modulated sinusoids, which specify the temporal envelopes, carrier frequencies, chirps, phases, and relative pulse shapes of the THz magnetic drive. The linear qAFM response is extracted from the oscillatory free-induction-decay component that remains in the transmitted single-pulse response after the main THz pulse has passed. Fitting this component to a damped sinusoid gives qAFM frequencies close to $0.61~\mathrm{THz}$, with decay times of $13.42~\mathrm{ps}$ and $12.61~\mathrm{ps}$ for the two pulses. The pulse and decay fits are summarized in Fig.~\ref{fig12}, and the fitting procedure is described in Appendix~\ref{app:experimental_fits}. In the experimental inference below, we use the averaged values $f_{\mathrm{qAFM}}=0.61~\mathrm{THz}$ and $\tau_{\mathrm d}=13.02~\mathrm{ps}$ as calibration targets: for each sampled Hamiltonian parameter set, $J$ is chosen to reproduce the qAFM frequency, while $\alpha$ is chosen to reproduce the qAFM mode decay time. In this way, the experimentally well-resolved linear response fixes the overall antiferromagnetic energy scale and damping, while the remaining Hamiltonian parameters are inferred from the nonlinear spectral information.

After fixing the pulse shapes and the linear qAFM calibration targets, the remaining experimental input needed for the forward simulations is the overall THz magnetic-field scale. The peak experimental electric field at the sample position is approximately $47.5~\mathrm{MV/m}$~\cite{Huang2024}, corresponding to a free-space magnetic-field amplitude $E^{(0)}/c\simeq0.16~\mathrm{T}$. To estimate the internal magnetic-field amplitude, we use a simple normal-incidence Fresnel correction,
\begin{align}
H_{\mathrm{int}}^{(0)}
\simeq
\frac{2n_{\mathrm{THz}}}{1+n_{\mathrm{THz}}}
\frac{E^{(0)}}{c}\,,
\end{align}
where $n_{\mathrm{THz}}$ is the THz refractive index of the sample. Using a typical rare-earth-orthoferrite THz refractive index $n_{\mathrm{THz}}\sim 4$--$5$~\cite{Zhou2012,Kim2015} gives an internal THz magnetic-field amplitude of order $0.25$--$0.27~\mathrm{T}$. We therefore use $H^{(0)}=0.27~\mathrm{T}$ as the magnetic-field scale in the experimental forward simulations. The proportionality factor relating the simulated nonlinear magnetization to the measured transmitted nonlinear THz electric field is not known a priori, because it depends on electromagnetic propagation through the sample, sample thickness, and the detection response. We therefore compare normalized spectra and time-domain traces below, so that the inference relies primarily on the relative spectral structure rather than on the absolute emitted or transmitted field amplitude.

With these forward-model inputs fixed, we generate the synthetic training libraries for the experimental inference task.
Both models are trained with $N_{\mathrm{samples}}=30000$ simulated spectra using the fitted experimental pulse shapes, the magnetic-field scale $H^{(0)}=0.27~\mathrm{T}$, $z$-polarized excitation with $\hat{\mathbf e}_1=\hat{\mathbf e}_2=\hat{\mathbf z}$, and the three-delay input $\tau=4.0$, $4.4$, and $4.8~\mathrm{ps}$. The input to the neural network is the standardized nonlinear $z$-component spectrum. From the experimental time-domain traces, we estimate an effective noise level of approximately $6\%$; the noise level is added to the synthetic time-domain nonlinear magnetization traces used for the corresponding noisy training and benchmark data. The expected inference accuracy under these experimental conditions is evaluated using independent noisy synthetic test sets with $N_{\mathrm{test}}=1000$ spectra. The reported error statistics are averaged over ten independently trained network seeds. For the two-parameter model, the relative MAEs are $14.7\pm0.4\%$ for $D$ and $49.9\pm1.1\%$ for $K_a$. For the three-parameter model, the corresponding relative MAEs are $17.2\pm0.5\%$ for $D$, $72.8\pm2.3\%$ for $K_a$, and $59.8\pm2.0\%$ for $K_4$. These relatively large errors are consistent with the robustness tests above: the experimental application uses only three delay traces and contains substantial noise, whereas the synthetic benchmarks showed that increasing the number of delays improves parameter recovery. Thus, the inferred value of $D$ is expected to be the most robust, while the anisotropy parameters should be interpreted with larger uncertainty.


Having benchmarked the two trained inverse models under noise conditions comparable to the experiment, we now apply the experimental-inference preprocessing pipeline to the measured nonlinear THz response. For this inference task, the synthetic training spectra and experimental spectra are processed identically. For each selected delay, the single-pulse reference responses are subtracted from the two-pulse response to obtain the nonlinear transmitted THz field. We then select the after-pulse detection window, apply the same Hann window used in the simulations, and compute the Fourier spectrum along the detection-time axis. For both the synthetic and experimental inputs, all spectra are normalized by the largest qAFM peak amplitude among the three delays included in the inference input. The same type of common normalization is applied to the corresponding time-domain nonlinear traces. This preserves the relative signal amplitudes between delays while reducing sensitivity to the unknown overall scale relating the measured transmitted THz field to the simulated nonlinear magnetization. After normalization, the spectra are logarithmically compressed and standardized using the mean and standard deviation obtained from the simulated training set. The resulting experimental spectra are therefore represented in the same standardized input space as the synthetic spectra used to train the neural network.

The standardized experimental spectra are passed through the trained networks to obtain the inferred Hamiltonian parameters, which are then inserted back into the spin-dynamics forward model for validation against the experimental spectra and time-domain traces. For the two-parameter model, the seed-averaged inferred parameters over ten independently trained networks are
$D_{\mathrm{pred}}=(1.61\pm0.12)\times10^{-1}~\mathrm{meV}$ and
$K_{a,\mathrm{pred}}=(1.6\pm3.0)\times10^{-3}~\mathrm{meV}$. For the three-parameter model, the corresponding values are
$D_{\mathrm{pred}}=(1.23\pm0.59)\times10^{-1}~\mathrm{meV}$,
$K_{a,\mathrm{pred}}=(1.6\pm2.3)\times10^{-3}~\mathrm{meV}$, and
$K_{4,\mathrm{pred}}=(2.3\pm5.2)\times10^{-5}~\mathrm{meV}$. The quoted uncertainties denote the seed-to-seed standard deviation over independently trained networks. Although the inferred parameters are restricted to the positive range used in training, the large standard deviations of $K_a$ and $K_4$ show that these parameters are only weakly constrained by the present experimental input.

Including $K_4$ as an additional inferred parameter changes the seed-averaged value of $D$ and increases its seed-to-seed spread, while the inferred scale of $K_a$ remains similar within the large uncertainty. The large relative uncertainties of $K_a$ and $K_4$ indicate that the experimental spectra constrain these anisotropy parameters less tightly than $D$ within the present reduced model and preprocessing pipeline. The comparison between the two- and three-parameter models should therefore be interpreted mainly as a test of whether including an additional nonlinear anisotropy improves the forward reconstruction, rather than as a definitive extraction of a unique quartic anisotropy value.

\begin{figure*}[t!]
\begin{center}
\includegraphics[scale=0.47]{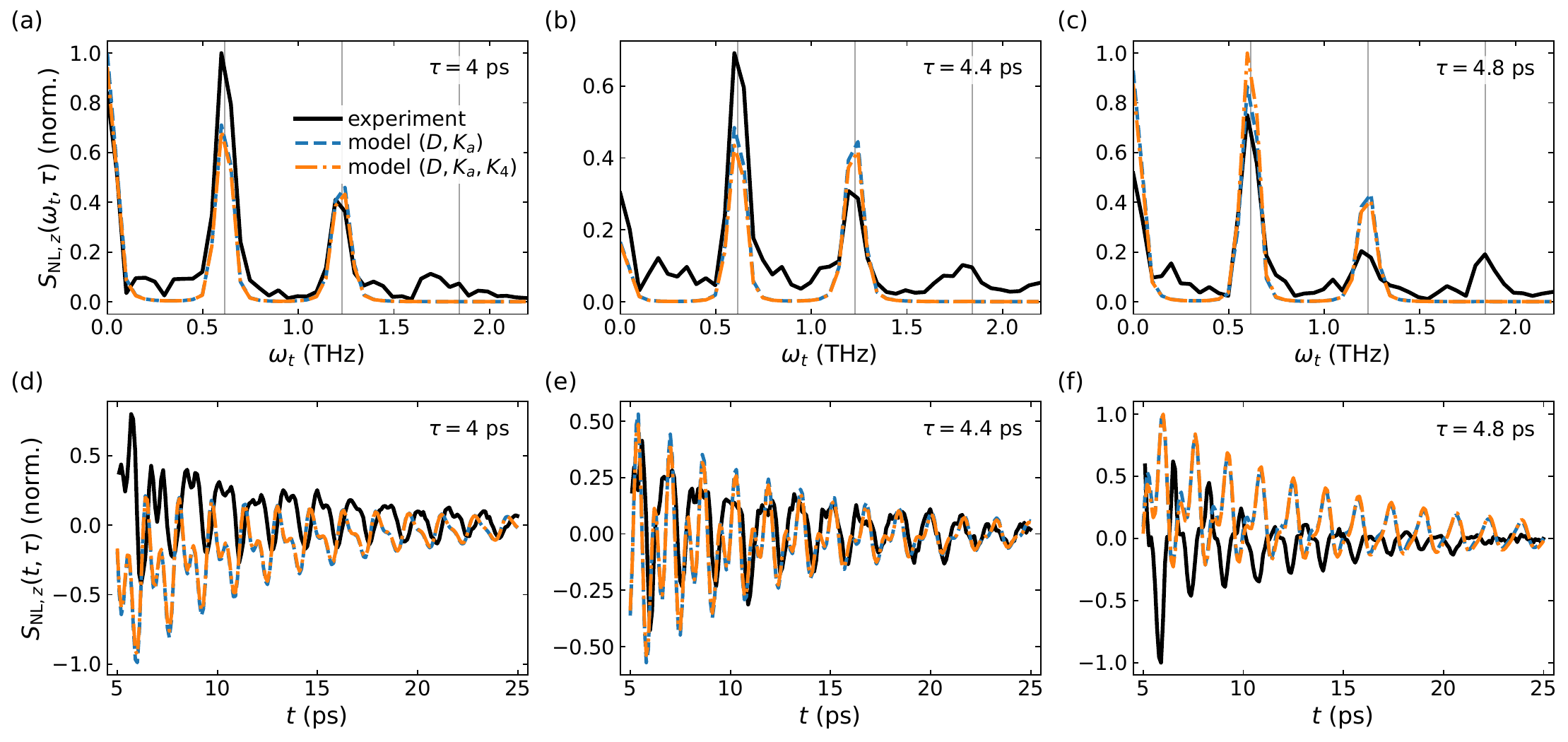}
\caption{
Experimental nonlinear response and forward simulations using inferred Hamiltonian parameters.
Comparison between the experimental nonlinear THz response of Sm$_{0.4}$Er$_{0.6}$FeO$_3$ and forward simulations obtained from two neural-network inference models. The first model infers $\bth=(D,K_a)$, while the second model infers $\bth=(D,K_a,K_4)$. Both models are trained with $N_{\mathrm{samples}}=30000$ simulated spectra using $z$-polarized excitation with $\hat{\mathbf e}_1=\hat{\mathbf e}_2=\hat{\mathbf z}$, the standardized nonlinear $z$-component spectral input, and an added time-domain noise level of $6\%$ estimated from the experimental traces. The models are evaluated using the three inter-pulse delays
$\mathcal T=\{4.0,4.4,4.8\}~\mathrm{ps}$. The forward simulations use the experimentally fitted THz pulse shapes scaled to the internal magnetic-field amplitude $H^{(0)}=0.27~\mathrm{T}$.
(a--c) Normalized nonlinear spectra for (a) $\tau=4.0~\mathrm{ps}$, (b) $\tau=4.4~\mathrm{ps}$, and (c) $\tau=4.8~\mathrm{ps}$. The experimental spectrum is shown by the black solid curve, while the blue dashed and orange dash-dotted curves show forward simulations obtained from the inferred $(D,K_a)$ and $(D,K_a,K_4)$ models, respectively. Each curve family is normalized by its maximum spectral amplitude across the displayed delays. Vertical lines mark integer multiples of the qAFM frequency $f_{\mathrm{qAFM}}=0.61~\mathrm{THz}$.
(d--f) Corresponding normalized after-pulse time-domain nonlinear responses for the same delays. The experimental traces and simulated $M_{\mathrm{NL},z}(t,\tau)$ responses are cropped to their respective after-pulse windows and normalized by the maximum time-domain amplitude of each curve family across the displayed delays.
The simulated curves correspond to one representative trained network seed. The seed-averaged inferred parameters over ten independently trained networks are
$D_{\mathrm{pred}}=(1.61\pm0.12)\times10^{-1}~\mathrm{meV}$ and
$K_{a,\mathrm{pred}}=(1.6\pm3.0)\times10^{-3}~\mathrm{meV}$ for the $(D,K_a)$ model, and
$D_{\mathrm{pred}}=(1.23\pm0.59)\times10^{-1}~\mathrm{meV}$,
$K_{a,\mathrm{pred}}=(1.6\pm2.3)\times10^{-3}~\mathrm{meV}$, and
$K_{4,\mathrm{pred}}=(2.3\pm5.2)\times10^{-5}~\mathrm{meV}$ for the $(D,K_a,K_4)$ model.
}
\label{fig11}
\end{center}
\end{figure*}

Figure~\ref{fig11} compares the experimental nonlinear response with forward simulations obtained from the inferred parameter sets. The upper row, Figs~\figref{fig11}{(a-c)}, shows normalized nonlinear spectra for (a) $\tau=4.0$, (b) $4.4$, and (c) $4.8~\mathrm{ps}$, while the lower row, Figs~\figref{fig11}{(d-f)}, shows the corresponding normalized after-pulse time-domain responses. The black curves show the experimental data, while the blue dashed and orange dash-dotted curves show simulations using the inferred parameters from the $(D,K_a)$ and $(D,K_a,K_4)$ models, respectively. The spectra and time-domain traces are normalized separately for each curve family by the maximum amplitude over the delays included in the inference input.

Because the qAFM frequency and damping are used as calibration targets for the forward model, agreement in the qAFM peak position and the overall oscillation/decay scale is expected and should not by itself be interpreted as an independent validation of the inferred parameters. However, the relative qAFM peak amplitudes across delays are not fixed by this calibration, because the spectra and time-domain traces are normalized only by a common maximum over the displayed delays. Differences in these amplitudes therefore remain meaningful. The more stringent test is whether the inferred parameters reproduce the relative nonlinear spectral structure, harmonic content, delay-dependent peak amplitudes, and time-domain phase beyond the calibrated frequency and damping scales. In this respect, the forward simulations capture part of the experimental nonlinear response: both inferred models generate nonlinear spectral weight in the relevant frequency range and reproduce some of the overall spectral structure. This shows that the learned inverse map produces physically reasonable parameter sets that can be inserted back into the spin-dynamics model for forward validation.

At the same time, the agreement is not quantitative across the full nonlinear response. This is consistent with the large seed-to-seed uncertainties of the experimentally inferred anisotropy parameters discussed above, which indicate that the present experimental input does not uniquely constrain all Hamiltonian parameters within the reduced two-sublattice model. The best spectral agreement is obtained for $\tau=4.0~\mathrm{ps}$, where the simulations reproduce the overall distribution of nonlinear spectral weight most closely. For the later delays, however, clear discrepancies remain. The most visible difference appears in the third-harmonic region, where the experimentally observed spectral feature is not well captured by either effective model. Including $K_4$ produces only minor changes in the simulated nonlinear spectrum and does not improve the agreement with the experimental third-harmonic peak. The limited improvement obtained by including $K_4$ does not contradict Ref.~\cite{huang2024extreme}, where a larger effective THz magnetic-field amplitude was used and the high-order nonlinear response was therefore more sensitive to the quartic anisotropy; at the experimentally calibrated field strength used here, the $K_4$ contribution is much weaker and does not by itself recover the missing third-harmonics peaks. In particular, neither the spectral weight nor the line shape of the observed third-harmonic feature is reproduced quantitatively. This suggests that the discrepancy is not primarily resolved by adding a local quartic anisotropy within the present two-sublattice Hamiltonian, but instead points to nonlinear mechanisms or modeling ingredients beyond this reduced description.

The time-domain comparison shows a similar limitation. The simulated after-pulse oscillations reproduce the general qAFM frequency and decay scale, but their phase does not consistently match the experimental response across the three delays. The phase agreement is best for $\tau=4.4~\mathrm{ps}$, while clear phase offsets remain for $\tau=4.0$ and $4.8~\mathrm{ps}$. Therefore, the experimental comparison should be interpreted as a proof-of-principle demonstration of the inference workflow on measured THz-2DCS data, rather than as a complete microscopic reconstruction of the measured nonlinear dynamics.

Several factors may contribute to these discrepancies. First, the forward model uses an effective two-sublattice description, whereas the real orthoferrite has a four-sublattice magnetic structure and may include additional exchange, anisotropy, or mode-coupling terms that are not included in the synthetic training library. Such terms may be especially important for higher-harmonic generation and could affect the experimentally observed third-harmonic response. 
Second, the present simulations treat the spin dynamics at the classical LLG level. Quantum spin effects, which are not included in this description, can modify nonlinear transition amplitudes and may enhance higher-order nonlinear responses compared with a purely classical treatment~\cite{mootz2023twodimensional}. This could also contribute to the underestimated third-harmonic spectral weight.
Third, the simulated observable is the nonlinear magnetization, while the experiment measures a transmitted THz electric field. The common normalization reduces sensitivity to the unknown overall conversion factor, but it does not remove possible frequency-dependent propagation effects, phase shifts, or detection transfer functions. In addition, the emitted nonlinear THz field can in general contain contributions from both magnetization and polarization dynamics~\cite{Srivastava2026PRB}. The present model uses the nonlinear magnetization alone as a proxy for the measured response and therefore neglects possible electric-dipole or magnetoelectric contributions, which may also affect the relative harmonic intensities.
Fourth, uncertainties in the pulse shape, polarization, sample alignment, and background subtraction can influence the relative amplitude and phase of the nonlinear time-domain traces. Finally, the experimental signal may contain contributions from weak modes, magnetic-domain averaging, or nonmagnetic nonlinearities that are absent from the reduced spin model.

Overall, Fig.~\ref{fig11} highlights both the promise and the current limitations of the approach. The inferred parameter sets generate forward simulations that reproduce the calibrated qAFM frequency scale and capture part of the nonlinear spectral structure observed experimentally. However, the remaining discrepancies in the third-harmonic spectral weight, line shape, and delay-dependent temporal phase show that a more complete forward model is needed for quantitative experimental Hamiltonian reconstruction. Natural extensions include incorporating the full four-sublattice orthoferrite structure, additional anisotropy and exchange terms, electromagnetic propagation and detection transfer functions, and training directly on simulated transmitted fields rather than only on nonlinear magnetization. Even with the present reduced model, the workflow demonstrates how experimental THz nonlinear spectra can be connected to microscopic Hamiltonian parameters through a calibrated synthetic training library and a forward-validation step.

\section{Conclusion and outlook}
\label{sec:conclusion}

We have introduced a supervised-learning framework for inferring effective Hamiltonian parameters from nonlinear THz-2DCS data. The central idea is to treat the nonlinear spectroscopic response as a high-dimensional fingerprint of the underlying Hamiltonian, rather than reducing it to a small number of manually selected peak positions, linewidths, or amplitudes. A calibrated forward model generates synthetic nonlinear spectra over a physically motivated parameter range, the simulated and experimental responses are processed through the same spectral pipeline, and a CNN learns the inverse map from nonlinear spectra to Hamiltonian parameters. The inferred parameters are then inserted back into the forward model, providing a direct validation step through comparison between predicted and reference nonlinear responses.

As a concrete demonstration, we applied this workflow to spin-Hamiltonian inference in a weakly canted antiferromagnet motivated by the room-temperature $\Gamma_4$ phase of Sm$_{0.4}$Er$_{0.6}$FeO$_3$. The effective two-sublattice model includes nearest-neighbor exchange, Dzyaloshinskii--Moriya interaction, quadratic anisotropy, quartic anisotropy, and Zeeman coupling to the THz magnetic field, while the spin dynamics are propagated using Landau–Lifshitz–Gilbert equations of motion with Gilbert damping. The exchange scale and damping are calibrated to the qAFM frequency and decay rate, while the remaining Hamiltonian parameters are inferred from nonlinear spectra. Synthetic benchmarks show that $z$-polarized excitation provides strong sensitivity to $D$, $K_a$, and $K_4$, but not to $K_c$ within the reduced model and $z$-component readout. In contrast, equal $y$--$z$-polarized excitation activates both qAFM and qFM dynamics and enables inference of $K_c$ as well. In the absence of added noise, the synthetic benchmarks show that the relevant Hamiltonian parameters can be inferred quantitatively from nonlinear spectra. The learning performance follows the physical sensitivity of the spectra: $D$ is generally recovered most accurately, while anisotropy parameters associated with subtler line-shape changes, sidebands, and higher-harmonic features are more challenging.

We also tested the robustness of the inverse model against time-domain noise and variations in the number of included inter-pulse delays. The results show that the delay axis of THz-2DCS provides important additional constraints. Multi-delay inputs reduce prediction errors, stabilize inferred parameters across independently trained networks, and improve forward prediction of noisy spectra and time-domain responses. The DM interaction $D$ remains the more robustly inferred parameter under noise, consistent with its stronger spectral fingerprint, whereas the anisotropy $K_a$ is more noise-sensitive and benefits particularly from the additional delay-dependent information. This demonstrates that Hamiltonian inference benefits not only from the spectral structure at a single delay, but also from the correlated evolution of nonlinear features with inter-pulse delay. Taken together with the parameter sweeps in which the qAFM frequency is held fixed, these results show that the information useful for Hamiltonian inference is not limited to the calibrated linear resonance frequency, but is distributed across nonlinear harmonics, polarization-dependent response channels, and the delay dependence of the THz-2DCS signal.

Finally, we applied the Hamiltonian-inference framework to experimental nonlinear THz data from Sm$_{0.4}$Er$_{0.6}$FeO$_3$. The inferred parameter sets produce forward simulations with physically reasonable nonlinear spectral features, but the comparison with experiment also reveals clear limitations of the present reduced model. In particular, the experimentally observed third-harmonic response is not fully captured, and the simulated after-pulse phase does not consistently match the experimental traces across all delays. Comparing the two-parameter $(D,K_a)$ and three-parameter $(D,K_a,K_4)$ models shows that including the quartic anisotropy produces only minor changes in the reconstructed response and does not resolve the discrepancy in the third-harmonic region. Moreover, the experimental data do not uniquely constrain this higher-order term within the present reduced model. These discrepancies indicate that the present implementation should be viewed as a proof-of-principle demonstration of the inference workflow on measured THz-2DCS data, rather than as a complete microscopic reconstruction of the measured nonlinear response.

Several extensions are natural next steps. On the modeling side, more quantitative comparison with experiment will require going beyond the effective two-sublattice approximation used here. Incorporating the full four-sublattice orthoferrite structure, additional exchange and anisotropy terms, weak mode couplings, and magnetic-domain effects would provide a more faithful spin-dynamics model. It will also be important to explore quantum-spin descriptions beyond the classical LLG approximation, which may modify higher-order spectral features~\cite{mootz2023twodimensional}. Finally, connecting the simulated response more directly to the experimentally measured transmitted THz field will require including electromagnetic propagation through the sample and detection transfer functions~\cite{Yang2021,Mikhaylovskiy2015}, rather than using the nonlinear magnetization alone as a proxy for the measured response.

On the data-analysis side, an important future direction is to use the full two-dimensional nonlinear spectrum for learning. In the present work, most benchmarks use one-dimensional detection-frequency spectra at selected delays, $S_{\mathrm{NL},\nu}(\omega_t,\tau)$, with different delays or magnetization components stacked as input channels. A more complete THz-2DCS representation would use the full delay-frequency response, for example $S_{\mathrm{NL}}(\omega_t,\omega_\tau)$, as the machine-learning input. Such an approach could exploit correlations between detection frequency and excitation-delay frequency, distinguish different nonlinear pathways more directly, and incorporate phase-sensitive information when available. Learning from full 2D spectra may therefore improve the inference of weak interaction terms and further stabilize the inverse problem against noise, extending the robustness gained from multi-delay inputs in the present benchmarks to the full delay-frequency response.

More broadly, the workflow developed here is not restricted to rare-earth orthoferrites or to spin systems. The same strategy can be applied to other driven quantum materials whenever a forward model can generate nonlinear multidimensional spectra over a relevant parameter range. By combining calibrated nonequilibrium simulations, multidimensional THz spectroscopy, and data-driven inverse modeling, this approach provides a route toward effective Hamiltonian reconstruction in regimes where linear spectroscopy alone is insufficient to separate competing microscopic interactions.

\section*{Data availability statement}
The data that support the findings of this study are openly available in figshare~\cite{Mootz2026data_MLspin}.


\section*{Acknowledgements}
This work was supported by the U.S. Department of Energy (DOE), Office of Science, Basic Energy Sciences, Materials Science and Engineering Division, including the grant of computer time at the National Energy Research Scientific Computing Center (NERSC) in Berkeley, California. The research was performed at the Ames National Laboratory, which is operated for the U.S. DOE by Iowa State University under Contract No. DE-AC02-07CH11358.



\appendix

\section{Construction and calibration of the forward model}
\label{app:forward_model}

This appendix summarizes how the calibrated forward model used in the main text is constructed. The Hamiltonian parameters inferred or sampled in the synthetic libraries are $\bth=(D,K_a,K_c,K_4)$. The exchange constant $J$ and Gilbert damping parameter $\alpha$ are not treated as independent learning targets. Instead, they are calibration parameters fixed by the linear qAFM response. For each sampled parameter set, $J$ is chosen such that the linearized qAFM frequency matches the prescribed target value $f_{\mathrm{qAFM}}$. After this frequency calibration, $\alpha$ is chosen such that the same qAFM mode reproduces the prescribed target decay rate $\Gamma_{\mathrm{target}}$. The parameter set used in the nonlinear time-domain simulation is therefore $(D,K_a,K_c,K_4;J,\alpha)$.

The calibration of $J$ is performed differently for $K_4=0$ and $K_4\neq0$. For $K_4=0$, both the equilibrium canting angle and the qAFM frequency are available in closed form. In this case, $J$ is obtained directly from the analytic qAFM frequency formula, and the equilibrium canting angle is then evaluated using the calibrated value of $J$. For $K_4\neq0$, the quartic anisotropy modifies both the equilibrium canting angle and the linearized mode frequency. In that case, $J$ is determined numerically: for each trial value of $J$, the equilibrium state is recomputed, the qAFM frequency is extracted from the linearized dynamics about that equilibrium state, and $J$ is adjusted until the target frequency is matched. Thus, the canting angle is not fixed independently before the calibration of $J$; instead, it is determined self-consistently from the calibrated Hamiltonian.

\subsection{Equilibrium configuration}
\label{app:equilibrium}

The equilibrium configuration is obtained within the canted two-sublattice manifold defined in Eq.~\eqref{eq:canted_ansatz}. The corresponding reduced energy $E(\varphi)$ and stationarity equation are given in Eqs.~\eqref{eq:reduced_energy_phi}--\eqref{eq:stationarity_phi}, with $\varphi=2\theta$. In the following, we discuss how these equations are solved during the calibration procedure.

For $K_4=0$, the stationarity equation has the analytic solution
\begin{align}
\tan\varphi
=
-\frac{M}{A}
=
-\frac{nD}{nJ+K_a-K_c}\,.
\label{eq:tanphi_appendix}
\end{align}
This equation gives two stationary solutions separated by $\pi$. Both candidates are evaluated in the reduced energy $E(\varphi)$, and the one with the lower energy is selected as the equilibrium branch. The canting angle used to construct the equilibrium spins is then
\begin{align}
\theta_{\mathrm{eq}}=\frac{\varphi_{\mathrm{eq}}}{2}\,.
\end{align}
In the $K_4=0$ calibration, this step is carried out after the analytic value of $J$ has been obtained from the qAFM frequency condition, as discussed in Sec.~\ref{app:J_calibration}.

For $K_4\neq0$, the quartic anisotropy changes the equilibrium angle, so the stationarity equation is solved numerically. For each trial value of $J$ used during the qAFM-frequency calibration, Eq.~\eqref{eq:stationarity_phi} is solved by a multi-start Newton procedure on the interval $[0,2\pi)$. Duplicate roots are removed, the reduced energy is evaluated for all stationary solutions, and the lowest-energy stationary point is selected. Thus, during the numerical calibration of $J$, the equilibrium angle is recomputed for every trial value of $J$ rather than being fixed in advance.

As a final consistency check, the selected spin configuration is inserted back into the full static Hamiltonian and the residual torque is evaluated,
\begin{align}
\tau_{\mathrm{res}}
=
\max_i
\left\lVert
\mathbf S_i\times\mathbf h_i^{\mathrm{eff}}
\right\rVert\,,
\qquad
\mathbf h_i^{\mathrm{eff}}
=
-\frac{\partial\mathcal H_0}{\partial\mathbf S_i}\,.
\end{align}
Here $\mathcal H_0$ is the static Hamiltonian without the THz driving term, and $\mathbf h_i^{\mathrm{eff}}$ is the effective field in energy units. Only states satisfying $\tau_{\mathrm{res}}<10^{-7}$ in units of $\mathrm{meV}\,|S|$ are retained.

\subsection{Calibration of $J$ from the qAFM frequency}
\label{app:J_calibration}

The exchange constant $J$ is chosen so that the linearized qAFM frequency agrees with the prescribed target value $f_{\mathrm{qAFM}}$. For $K_4=0$, the qAFM frequency can be written analytically as~\cite{Zhang2024Down}
\begin{align}
f_{\mathrm{qAFM}}
=
\frac{S}{2\pi\hbar}
\left[
4K_a\left(nJ+K_a-K_c\right)
+
(nD)^2
\right]^{1/2}\,.
\label{eq:qafm_closed_form_app}
\end{align}
Solving Eq.~\eqref{eq:qafm_closed_form_app} for $J$ gives
\begin{align}
J
=
\frac{1}{n}
\left[
\frac{
\left(2\pi\hbar f_{\mathrm{qAFM}}/S\right)^2
-
(nD)^2
}{4K_a}
-
(K_a-K_c)
\right]\,.
\label{eq:J_from_qafm_app}
\end{align}
This expression is used for the $K_4=0$ calibration, with $f_{\mathrm{qAFM}}$ set to the target qAFM frequency. Parameter sets are retained only if the calibrated exchange constant satisfies
\begin{align}
0<J\le J_{\max}=10~\mathrm{meV}
\end{align}
and the final equilibrium configuration passes the torque criterion above; all other parameter sets are discarded. After $J$ has been obtained from Eq.~\eqref{eq:J_from_qafm_app}, the equilibrium canting angle is evaluated from Eq.~\eqref{eq:tanphi_appendix} using this calibrated value of $J$.

For $K_4\neq0$, Eq.~\eqref{eq:J_from_qafm_app} can no longer be used  because the quartic anisotropy changes both the equilibrium canting angle and the linearized qAFM frequency. Instead, $J$ is found by a numerical root search. For each trial value of $J$, the equilibrium spin configuration $\{\mathbf S_i^{(0)}\}$ is first determined self-consistently using the procedure described in Sec.~\ref{app:equilibrium}. The undamped, field-free equations of motion are then linearized about this equilibrium state.

To obtain the linear normal modes, we use the standard linearization of fixed-length Landau--Lifshitz spin dynamics in the tangent plane of the equilibrium spin configuration~\cite{Zhuonan2023}. To preserve the fixed spin length to linear order, the fluctuation of each spin is restricted to the plane perpendicular to the corresponding equilibrium spin,
\begin{align}
\mathbf S_i(t)
=
\mathbf S_i^{(0)}
+
\delta\mathbf S_i(t)\,,
\qquad
\delta\mathbf S_i(t)\cdot\mathbf S_i^{(0)}=0\,.
\end{align}
This follows from
\begin{align}
\left|\mathbf S_i^{(0)}+\delta\mathbf S_i\right|^2
=
S^2
+
2\mathbf S_i^{(0)}\cdot\delta\mathbf S_i
+
\mathcal O(\delta\mathbf S_i^2)\,,
\end{align}
so transverse fluctuations preserve the spin length to first order and describe only changes of spin direction.

For each spin, the deviation is expanded in two orthonormal directions spanning the tangent plane perpendicular to the equilibrium spin,
\begin{align}
\delta\mathbf S_i(t)
=
x_{i1}(t)\mathbf e_{i1}
+
x_{i2}(t)\mathbf e_{i2}\,,
\end{align}
with
\begin{align}
\mathbf e_{i1}\cdot\mathbf S_i^{(0)}
=
\mathbf e_{i2}\cdot\mathbf S_i^{(0)}
=
0\,,
\qquad
\mathbf e_{i\mu}\cdot\mathbf e_{i\nu}
=
\delta_{\mu\nu}\,.
\end{align}
The particular choice of transverse basis does not affect the resulting mode frequencies. The four fluctuation amplitudes are collected into
\begin{align}
\mathbf x
=
(x_{11},x_{12},x_{21},x_{22})^T\,.
\end{align}

For the undamped, field-free dynamics, we define the right-hand side
\begin{align}
\mathbf F_i(\{\mathbf S_j\})
=
-\frac{1}{\hbar}\,
\mathbf S_i\times\mathbf h_i^{\mathrm{eff}}\,,
\qquad
\mathbf h_i^{\mathrm{eff}}
=
-\frac{\partial\mathcal H_0}{\partial\mathbf S_i}\,.
\label{eq:Fi_undamped_app}
\end{align}
Keeping only terms linear in the fluctuation amplitudes gives
\begin{align}
\frac{d\mathbf x}{dt}
=
\mathcal J\mathbf x\,,
\end{align}
where the projected Jacobian has matrix elements
\begin{align}
\mathcal J_{ia,jb}
=
\mathbf e_{ia}\cdot
\left.
\frac{\partial \mathbf F_i}{\partial x_{jb}}
\right|_{\mathbf x=0}\,.
\label{eq:projected_jacobian_app}
\end{align}
In the numerical implementation, these derivatives are evaluated by finite differences. Each spin is displaced by a small amount along one of its transverse basis directions, renormalized to length $S$, and the resulting change in $\mathbf F_i$ is projected back onto the four transverse basis directions.

Diagonalizing $\mathcal J$ gives the small-amplitude normal-mode frequencies. For the undamped dynamics considered here, the eigenvalues occur in pairs $\lambda=\pm i\Omega_\nu$. The positive-frequency mode whose ordinary frequency $\Omega_\nu/(2\pi)$ is closest to the target value $f_{\mathrm{qAFM}}$ is identified as the qAFM branch, while the lower positive-frequency mode is identified as the qFM branch. The trial value of $J$ is then updated by bisection until the computed qAFM frequency agrees with the target value $f_{\mathrm{qAFM}}$. Parameter sets for which the target frequency cannot be reached within the accepted interval $0<J\le J_{\max}=10~\mathrm{meV}$ are discarded.

\subsection{Calibration of the damping parameter $\alpha$}
\label{app:alpha_calibration}

After $J$ has been fixed, the Gilbert damping parameter $\alpha$ is determined from the target decay rate of the qAFM mode. The static equilibrium configuration does not depend on $\alpha$, so the equilibrium state obtained after the $J$ calibration is kept fixed during the damping calibration.

For a trial value of $\alpha$, the linearized damped dynamics are constructed using the same transverse fluctuation coordinates introduced in Sec.~\ref{app:J_calibration}. The undamped vector field $\mathbf F_i$ is replaced by the field-free damped LLG right-hand side,
\begin{align}
\mathbf F_i^{(\alpha)}(\{\mathbf S_j\})
=
-\frac{1}{\hbar(1+\alpha^2)}
\left[
\mathbf S_i\times\mathbf h_i^{\mathrm{eff}}
+
\frac{\alpha}{S}
\mathbf S_i\times
\left(
\mathbf S_i\times\mathbf h_i^{\mathrm{eff}}
\right)
\right]\,,
\label{eq:Fi_damped_app}
\end{align}
where $\mathbf h_i^{\mathrm{eff}}$ is evaluated from the static Hamiltonian, as in Eq.~\eqref{eq:Fi_undamped_app}. The corresponding damped linearized matrix is obtained by the same finite-difference projection used in Eq.~\eqref{eq:projected_jacobian_app}, but with $\mathbf F_i$ replaced by $\mathbf F_i^{(\alpha)}$.

\begin{figure*}[t!]
\begin{center}
\includegraphics[scale=0.78]{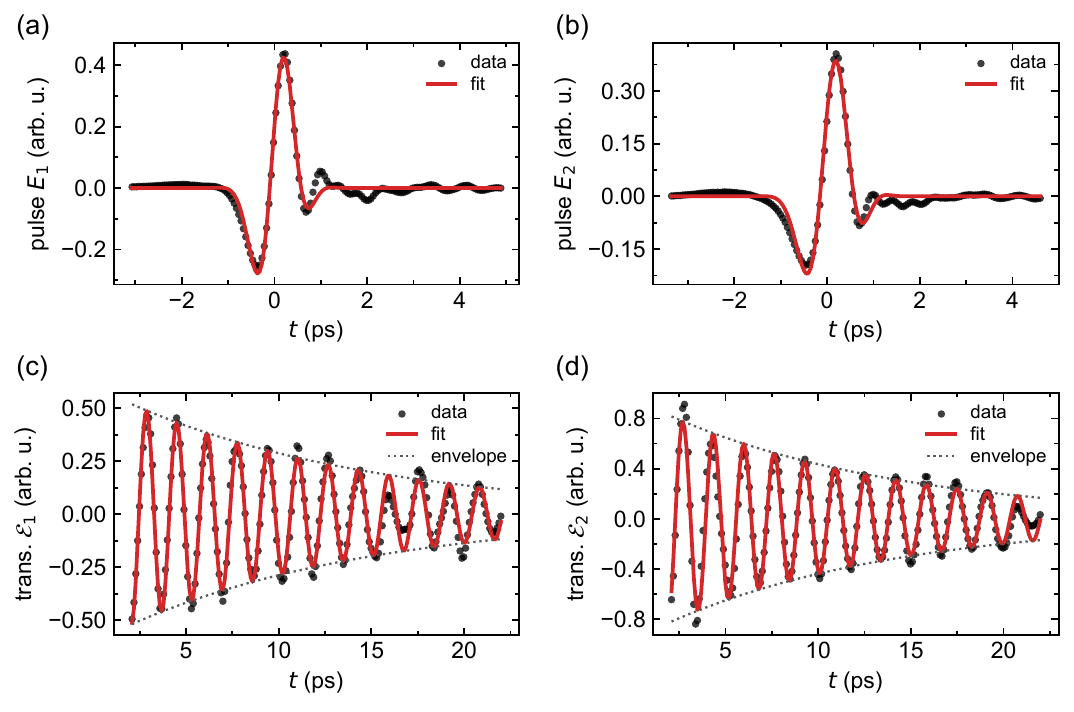}
\caption{
Pulse and decay fits used to characterize the THz excitation and transmitted qAFM response.
(a,b) Experimental incident THz pulse waveforms $E_1$ and $E_2$ (markers), together with fits to a chirped Gaussian-modulated sinusoid, Eq.~\eqref{eq:chirped_pulse} (solid lines).
(c,d) Experimental transmitted-field traces $\mathcal E_1$ and $\mathcal E_2$ following excitation by $E_1$ and $E_2$ (markers), together with fits to a damped oscillation, Eq.~\eqref{eq:decay_fit_app} (solid lines). The dotted lines in (c,d) indicate the fitted decay envelopes.
}
\label{fig12}
\end{center}
\end{figure*}

Diagonalizing the damped matrix gives eigenvalues $\lambda$ defined by the time dependence
$\mathbf x(t)\propto e^{\lambda t}$. Since the linearized matrix is real, each damped oscillatory mode appears as a complex-conjugate pair,
\begin{align}
\lambda_{\nu,\pm}
=
-\Gamma_\nu \pm i\,2\pi f_\nu^{\mathrm{lin}}\,,
\end{align}
where $\nu$ labels the normal-mode branch. The two eigenvalues in each pair describe the same real damped oscillation, with opposite signs of the oscillation frequency. We use the positive-frequency representative, $\mathrm{Im}\,\lambda_{\nu,+}>0$, for each mode. Among the positive-frequency modes, the qAFM branch is identified as the mode whose frequency $\mathrm{Im}\,\lambda/(2\pi)$ is closest to the undamped qAFM frequency obtained after the $J$ calibration. For this branch, we write
\begin{align}
\lambda_{\mathrm{qAFM}}(\alpha)
=
-\Gamma_{\mathrm{qAFM}}(\alpha)
+
i\,2\pi f_{\mathrm{qAFM}}^{\mathrm{lin}}(\alpha)\,.
\label{eq:lambda_appendix}
\end{align}
Here $f_{\mathrm{qAFM}}^{\mathrm{lin}}(\alpha)$ is the ordinary frequency of the linearized qAFM mode, and $\Gamma_{\mathrm{qAFM}}(\alpha)$ is its decay rate. With this convention,
\begin{align}
f_{\mathrm{qAFM}}^{\mathrm{lin}}(\alpha)
=
\frac{1}{2\pi}
\mathrm{Im}\,[\lambda_{\mathrm{qAFM}}(\alpha)]\,,
\end{align}
and
\begin{align}
\Gamma_{\mathrm{qAFM}}(\alpha)
=
-\mathrm{Re}\,[\lambda_{\mathrm{qAFM}}(\alpha)]\,.
\end{align}

The Gilbert damping parameter is then obtained by solving
$\Gamma_{\mathrm{qAFM}}(\alpha)
=
\Gamma_{\mathrm{target}}$
by bisection in the interval $0\leq\alpha\leq5\times10^{-3}$. If the target decay rate cannot be reached within this interval, the sampled Hamiltonian parameter set is discarded. The final calibrated parameter set $(D,K_a,K_c,K_4;J,\alpha)$ is then used in the nonlinear time-domain simulations.

\section{Experimental pulse and decay fits}
\label{app:experimental_fits}

This appendix summarizes the fits used to characterize the experimental THz excitation pulses and the linear qAFM response used for model calibration. Figures~\figref{fig12}{(a,b)} show the measured THz excitation pulses $E_1$ and $E_2$ as markers, together with fits to the chirped Gaussian-modulated sinusoid introduced in Eq.~\eqref{eq:chirped_pulse} as solid lines. For these fits, the same functional form as Eq.~\eqref{eq:chirped_pulse} is used for the scalar electric-field waveform, with $t$ replaced by $t-t_0$ to allow for a temporal offset. The fitted pulse parameters are summarized in Table~\ref{tab:pulse_fit_params}. These fits provide the temporal pulse profiles used to construct the experimental drive in the forward simulations.

\begin{table}[H]
\centering
\caption{
Fitted parameters of the experimental incident THz pulse waveforms. The amplitude $E_0$ is given in arbitrary units.
}
\label{tab:pulse_fit_params}
\begin{tabular}{c c c c c c c}
\hline\hline
Pulse & $E_0$ & $\sigma$ (ps) & $f$ (THz) & $t_0$ (ps) & $\beta$ (ps$^{-1}$) & $\phi$ (rad) \\
\hline
$E_1$ & 0.50 & 0.56 & 0.68 & 0.12 & 0.20 & 0.42 \\
$E_2$ & 0.43 & 0.62 & 0.63 & 0.41 & 0.20 & 0.59 \\
\hline\hline
\end{tabular}
\end{table}

Figures~\figref{fig12}{(c,d)} show the transmitted single-pulse responses $\mathcal E_1$ and $\mathcal E_2$ induced by excitation pulses $E_1$ and $E_2$ in the after-pulse window as markers, together with fits to a damped free-induction-decay (FID) oscillation as solid lines,
\begin{align}
\mathcal E_{\mathrm{FID}}(t)
=
A
\exp\left(-\frac{t-t_0}{\tau_{\mathrm d}}\right)
\cos\left[
2\pi f(t-t_0)+\phi
\right]
+
b_0\,.
\label{eq:decay_fit_app}
\end{align}
Here $f$ is the qAFM oscillation frequency and $\tau_{\mathrm d}$ is the decay time. The fitted decay parameters are summarized in Table~\ref{tab:decay_fit_params}.

\begin{table}[H]
\centering
\small
\setlength{\tabcolsep}{3.5pt}
\caption{
Fitted parameters of the transmitted qAFM free-induction-decay response. The amplitude $A$ and offset $b_0$ are given in the arbitrary units of the transmitted-field traces.
}
\label{tab:decay_fit_params}
\begin{tabular}{c c c c c c c}
\hline\hline
Exc. & $A$ & $\tau_{\mathrm d}$ & $f$ & $t_0$ & $\phi$ & $b_0$ ($10^{-3}$) \\
& & (ps) & (THz) & (ps) & (rad) & \\
\hline
$E_1$ & $-0.51$ & 13.42 & 0.612 & 40.23 & 0.68 & 0.265 \\
$E_2$ & $-0.80$ & 12.61 & 0.609 & 40.29 & 1.54 & $-1.55$ \\
\hline\hline
\end{tabular}
\end{table}

The fitted frequencies confirm that the dominant after-pulse oscillation corresponds to the qAFM mode near $0.61~\mathrm{THz}$. In the calibrated forward model, this frequency fixes the exchange scale $J$ through the linearized qAFM mode. The fitted decay times provide the corresponding damping scale. Equivalently, the exponential decay rate is $\Gamma=1/\tau_{\mathrm d}$ for a time dependence $\exp(-\Gamma t)$. In the experimental inference, we use the averaged values $f_{\mathrm{qAFM}}=0.61~\mathrm{THz}$ and $\tau_{\mathrm d}=13.02~\mathrm{ps}$ as fixed experimental inputs $\boldsymbol{\eta}$ rather than as learning targets.


\bibliography{ref}

\end{document}